\documentclass[journal=jctc,manuscript=article]{achemso}

\SectionNumbersOn

\usepackage{amsmath,amsfonts,amssymb,amsthm}
\usepackage{xcolor}

\newcommand{\R}{\mathbb{R}}                                     

\newcommand{\pd}[2]{\frac{\partial#1}{\partial#2}}              
\newcommand{\td}[2]{\frac{\mathrm{d}#1}{\mathrm{d}#2}}          

\newcommand{\innerprod}[2]{\left\langle #1,\, #2 \right\rangle} 

\newcommand{\diff}{\mathrm{d}}                                   

\newcommand{\cL}{\mathcal{L}}

\newcommand{\BA}{\mathbf{A}}

\newcommand{\BM}{\mathbf{M}}

\newcommand{\BS}{\mathbf{S}}
\newcommand{\BT}{\mathbf{T}}
\newcommand{\BU}{\mathbf{U}}
\newcommand{\BV}{\mathbf{V}}
\newcommand{\BW}{\mathbf{W}}
\newcommand{\BX}{\mathbf{X}}

\newcommand{\BZ}{\mathbf{Z}}

\newcommand{\bbf}{\mathbf{f}}

\newcommand{\bq}{\mathbf{q}}

\newcommand{\bt}{\mathbf{t}}

\newcommand{\bw}{\mathbf{w}}
\newcommand{\bx}{\mathbf{x}}

\newcommand{\bz}{\mathbf{z}}

\title{Tensor-based Approximation of Molecular Kinetics: Generator Learning, Reaction Coordinates and Incremental Updating}
\author{Minakshi Verma}
\affiliation{ 
Otto-von-Guericke Universität, Faculty of Mathematics, Magdeburg 39106, Germany
}
\alsoaffiliation{ 
Max-Planck-Institute for Dynamics of Complex Technical Systems, Magdeburg 39106, Germany
}
\author{Peter Benner}
\affiliation{ 
Max-Planck-Institute for Dynamics of Complex Technical Systems, Magdeburg 39106, Germany
}
\author{Feliks Nüske}
\affiliation{ 
Max-Planck-Institute for Dynamics of Complex Technical Systems, Magdeburg 39106, Germany
}
\email{nueske@mpi-magdeburg.mpg.de}

\usepackage{graphicx}
\usepackage{booktabs}
\usepackage{xcolor}
\newcommand{\hl}[1]{%
  \ifmmode
    \textcolor{blue!70!black}{\mathbf{#1}}%
  \else
    \textcolor{blue!70!black}{\textbf{#1}}%
  \fi
}
\usepackage{amsmath}
\usepackage{amsfonts}
\usepackage{amsthm}
\usepackage{subcaption}
\usepackage{placeins}
\usepackage{stmaryrd}
\usepackage{mathtools}
\usepackage{float}
\usepackage{natbib}        
\usepackage{algorithm}
\usepackage{algpseudocode}
\usepackage[hidelinks]{hyperref}
\usepackage{graphicx}
\usepackage{dsfont}

\newcommand{\algrule}[1][.2pt]{\par\vskip.5\baselineskip\hrule height #1\par\vskip.5\baselineskip}

\begin{document}
\maketitle

\begin{abstract}
    We present an approach to analyze long-timescale kinetics of molecular dynamics simulations - meta-stable states and transition timescales - by a tensor-based approximation of the infinitesimal generator. We start from the variational approximation of low-lying generator eigenvalues, and discretize the associated eigenvalue problem using a tensor-product basis. We derive analytical expressions for the resulting tensor operators in tensor train (TT) format, and provide efficient algorithms to solve the corresponding linear problems. We also treat the case of a state-dependent diffusion field, which is relevant when working in reaction coordinates. We show how the number of reaction coordinates can be gradually increased without having to solve the variational problem from scratch. Using MD simulations of fast-folding proteins and TICA coordinates as reaction coordinates, we demonstrate the effectiveness of the proposed method at identifying long-timescale transitions and meta-stable states.
\end{abstract}

\section{Introduction}
Molecular Dynamics (MD)~\cite{frenkelUnderstandingMolecularSimulation2023,tuckermanStatisticalMechanicsTheory2023a} provides high-resolution simulations of the dynamical behaviour of large-scale molecular systems, with applications in biology, chemistry, materials science, and many other disciplines. While advances in hardware and software implementation have allowed to study larger and larger systems, a wide range of post-processing methods has emerged to estimate metastable states, transition timescales and mechanisms, and many other dynamical observables. 


Markov State Models (MSMs) are a particularly successful post-processing method. Based on Ulam's method~\cite{ulam_collection_1960}, MSMs were developed in the late 1990s~\cite{dellnitz_approximation_1999,schutte_direct_1999} and have enabled detailed studies of the conformational states and transition dynamics of complex systems, see Refs~\cite{noeConstructingEquilibriumEnsemble2009,plattnerCompleteProteinProtein2017b,knoverekOpeningCrypticPocket2021} for examples. The MSM approach was generalized by the variational approach to conformational dynamics (VAC)~\cite{noe_variational_2013,nuske_variational_2014}, which enabled the use of general non-linear basis functions instead of set-discretizations, and provided a rigorously defined scalar objective function to assess model quality. A fully general extension to non-reversible systems was presented by Wu and Noé in~\cite{wuVariationalApproachLearning2020c}. 

MSMs and the VAC are specific instances of approximation methods for the Koopman operator. Dating back to Koopman's seminal work from the 1930s~\cite{koopmanHamiltonianSystemsTransformation1931b}, and building on works by Mezic and co-workers~\cite{mezicSpectralPropertiesDynamical2005a}, Koopman modeling has emerged as a powerful machine-learning based framework for complex dynamical systems~\cite{budisicAppliedKoopmanism2012b,mauroyKoopmanOperatorSystems2020c,bruntonModernKoopmanTheory2022,colbrookChapter4Multiverse2024}. Both the VAC and MSMs are special cases of the \emph{extended dynamic mode decomposition} (EDMD) algorithm~\cite{williams_datadriven_2015} for Koopman operator learning, albeit derived from different principles~\cite{klus_data-driven_2018}.

A widely-used strategy for VAC / MSM modeling is to pass raw molecular features through an initial linear dimensionality reduction, often using time-independent component analysis (TICA)~\cite{molgedey_separation_1994,perez-hernandez_identification_2013}. On the resulting space of TICA coordinates (TICs), non-linear VAC models can then be trained to high precision~\cite{bowman_introduction_2014}. Non-linear models based on deep learning or kernel methods~\cite{schwantesModelingMolecularKinetics2015b,mardt_vampnets_2018,klusKernelbasedApproachMolecular2018b,chenNonlinearDiscoverySlow2019b,sprink_accurate_2026} have been demonstrated to be highly accurate, and allow for ease of implementation and validation. 

Tensor-product models have been developed in parallel~\cite{klus_tensor-based_2018,nuske_tensor-based_2021,cao_amuset-tica_2025}, and can achieve similar precision. Tensor techniques have gained significant momentum in computational mathematics, sciences and engineering, due to their ability to compress high-dimensional data sets by exploiting low-rank structures. The starting point here is the Canonical Polyadic Decomposition (CPD), also known as CANDECOMP/PARAFAC, which truncates the sum of rank-1 tensors needed to express a multidimensional array. Other formats that are more useful in practice are the (hierarchical) Tucker format, and in particular its instance Tensor-Train (TT) decomposition, which is known in computational physics as Matrix Product States (MPS) and will be the format used here. For more details, precise definitions, and algorithms to compute the various tensor decompositions, see, in particular, the monographs \cite{BalK2025,Kho2018, KhoK2018} and references therein. Tensor-based models are generally more intricate in terms of implementation and model validation, but are in a sense more structured, as one can more directly assess the contribution of individual molecular degrees of freedom on the final non-linear model. One of the goals of this paper is to provide a principled approach to make this assessment in an efficient manner.

Another line of research focuses on learning of the infinitesimal generator or Koopman generator of the molecular dynamics~\cite{klus_data-driven_2020,lucke_tgedmd_2022,zhang_solving_2022,devergne_learning_2024}, which is the evolution operator for the underlying Kolmogorov equation. The generator provides access to the same dynamical properties as the Koopman operator, but does not strictly require time-series data for training~\cite{devergne_slow_2025,moqvist_thermodynamic_2025,nateghi_consistent_2026}. In addition, the generator is less of a black-box, as it is intimately related to the parameters of the dynamics, which can be exploited to design learning methods~\cite{donati_estimation_2018}. The generator also encodes dynamical properties of coarse-grained models on reaction coordinates. This connection has been explored in detail in Refs.~\cite{zhangEffectiveDynamicsGiven2016a,zhang_reliable_2017,nateghi_kinetically_2025}.

In this study, we combine generator learning with tensor-based models in the VAC framework. The main contributions of our work can be summarized as follows:
\begin{itemize}
    \item we derive TT representations of the matrices required by the variational approach to conformational dynamics (VAC) for the reversible Koopman generator. This includes systems with state-dependent diffusion, which is the generic situation when operating on reaction coordinates.
    \item we develop optimized TT contraction procedures to efficiently set up a small-scale VAC matrix eigenvalue problem to compute transition rates and timescales.
    \item we develop an updating scheme which allows to iteratively screen an increasing number of reaction coordinates, such as an increasing number of TICs.
    \item we demonstrate that the tensor-based approach reproduces the dominant spectral quantities and metastable structures for model systems and MD simulations of fast-folding proteins.
\end{itemize}

Section~\ref{sec: Preliminaries} introduces the required preliminaries on the variational approach and generator approximation. In Section~\ref{sec:tensor_reps}, we introduce product basis functions and TT representations. 
Section~\ref{sec:main_algorithm} presents the key theoretical and algorithmic results of this study. In Section~\ref{sec:Applications}, we illustrate the performance of the method using a low-dimensional toy system and fast-folding protein simulations, including eigenvalue comparisons, rank behavior under truncation, and PCCA-based metastability analysis. 
Finally, Section~\ref{sec:Conclusion and Outlook} concludes the paper and discusses possible directions for future work.

\section{Preliminaries}
\label{sec: Preliminaries}

\subsection{Molecular Dynamics and Koopman Generator}
We model molecular dynamics as a stochastic process on Euclidean state space $\R^d$ governed by a reversible stochastic differential equation (SDE). A prototype dynamical model is the \emph{Langevin dynamics}
\begin{equation}
\label{eq:langevin_sde}
    \diff \bx_t = -\nabla V(\bx_t)\,\diff t + \sqrt{2 \beta^{-1}}\,\diff \bw_t,
\end{equation}
driven by the negative gradient of the potential energy $V$, with constant and isotropic diffusion at \emph{inverse temperature} $\beta$. The general form of a reversible SDE would be
\begin{equation}
\label{eq:sde}
    \diff \bx_t = \bbf(\bx_t)\,\diff t + \sqrt{2\beta^{-1}}\sigma(\bx_t)\,\diff \bw_t,
\end{equation}
with drift field $\bbf$, diffusion field $\sigma$, and invariant (Boltzmann) distribution
\begin{equation*}
    \diff \mu(\bx) = \frac{1}{Z_\beta} \exp(-\beta V(\bx))\,\diff \bx.
\end{equation*}
We reserve the symbol $\Sigma$ for its \emph{diffusion co-variance matrix}:
\begin{equation*}
    \Sigma(\bx) := (\sigma\sigma^\top)(\bx).
\end{equation*}

The statistics of the dynamics~\eqref{eq:sde} are governed by the \emph{Koopman generator} or infinitesimal generator (see e.g.~\cite{pavliotis_stochastic_2014}):
\begin{equation}
    \label{eq:generator}
    \cL \psi(\bx) = \bbf(\bx) \cdot \nabla \psi(\bx) + \frac{1}{\beta} \Sigma(\bx) : \nabla^2 \psi(\bx),
\end{equation}
which is a linear second-order differential operator with coefficients determined by the SDE~\eqref{eq:sde}. For Langevin dynamics, the generator $\cL$ reduces to 
\begin{equation*}
    \cL\psi(\bx) = -\nabla V(\bx) \cdot \nabla \psi(\bx) + \frac{1}{\beta} \Delta \psi(\bx).
\end{equation*}
Long-time properties of the dynamics are encapsulated in the low-lying spectrum of the negative generator $-\cL$: meta-stable states and the rates of transition in between are encoded by solutions of the eigenvalue equation
\begin{equation}
    \label{eq:ev_problem_generator}
    -\cL \varphi_i = \kappa_i \varphi_i,
\end{equation}
where $0 = \kappa_0 < \kappa_1 \leq \kappa_2 \ldots$ are the smallest eigenvalues, and $\varphi_i$ the associated eigenfunctions. The eigenvalues $\kappa_i$ can be interpreted as relaxation rates, whereas the eigenfunctions can be used to identify meta-stable states using spectral clustering methods, such as PCCA~\cite{deuflhardRobustPerronCluster2005a}.

For later use, we note the following important \emph{integration-by-parts} formula for reversible systems, which reflects the symmetry of the Koopman generator w.r.t. the inner product defined by $\mu$:
\begin{equation}
\label{eq:integration_by_parts}
    \innerprod{\psi_1}{\cL \psi_2}_\mu = -\frac{1}{\beta} \int_{\R^d} \nabla \psi_1(\bx)\cdot \Sigma(\bx) \cdot \nabla \psi_2(\bx)\,\diff \mu(\bx) = \innerprod{\cL\psi_1}{\psi_2}_\mu.
\end{equation}

\subsection{Variational Approach}
To compute the eigenvalues and eigenfunctions in~\eqref{eq:ev_problem_generator} numerically, we need to approximate this equation by a matrix eigenvalue problem. A principled approach to doing so is the \emph{variational approach to conformational dynamics (VAC)}~\cite{noe_variational_2013,nuske_variational_2014,zhang_reliable_2017,zhang_solving_2022}, which is equivalent to generator extended dynamic mode decomposition (gEDMD)~\cite{klus_data-driven_2020} for a reversible system. The procedure is summarized in Algorithm~\ref{alg:vac}.

\begin{figure}
    \begin{algorithm}[H]
    \caption{Variational Approximation of Generator Eigenpairs (gEDMD)}\label{alg:vac}
    \begin{algorithmic}[1]
    \Require Basis set $\Psi(\bx) = (\psi_1(\bx),\ldots,\psi_N(\bx))^\top$; training data $\bx_1,\ldots, \bx_m$ sampled according to the invariant measure $\mu$    
    \Ensure Approximate eigenvalues $\kappa_i$ and eigenfunctions $\varphi_i$ as solution of~\eqref{eq:ev_problem_generator}
    \algrule
    \State Set up data matrix $\Psi_\BX = \left[\psi_i(\bx_l) \right]_{i,l} \in \R^{n \times m}$ and gradient tensor $\nabla \Psi_\BX = \left[\pd{}{\bx^j}\psi_i(\bx_l) \right]_{i,j,l} \in \R^{n \times d \times m}$.
    \State Rank-$r$ SVD of the data matrix: $\frac{1}{\sqrt{m}}\Psi_\BX
\approx
\BU \Sigma \BV^\top$. \label{algline:whitening}
    \State Mean vector of reduced basis: $\bar{\bz} = \frac{1}{m} \BV^\top \mathbf{1}_m \in \mathbb{R}^{r}$.\label{algline:mean_vector}
    \State Mean-free correction to whitening transform: diagonalize $\mathrm{Id}_r  - \bar{\bz}\bar{\bz}^{\top} = \BW \,\mathrm{diag}(\lambda)\, \BW^{\top}$ and set $\BZ = \BW \mathrm{diag}(\lambda^{-1/2})$.\label{algline:mean_removal}
    \State Compute generator stiffness matrix: $\BA_{ij} = \frac{1}{\beta m}\sum_{l=1}^{m} \nabla \psi_i(\bx_l) \cdot \Sigma(\bx_l) \cdot \nabla \psi_j(\bx_l)$.\label{algline:stiffness_matrix}
    \State Reduced generator matrix: $\BM = \BZ^\top \Sigma^{-1} \BU^\top
\BA \BU \Sigma^{-1} \BZ \in \R^{r \times r}$.\label{algline:reduced_matrix}
    \State Small-scale eigenvalue problem: $\BM \bw_i = \kappa_i \bw_i$.
    \State Evaluate eigenfunctions: $\varphi_i(\bx) = \Psi(\bx)^\top \BU \Sigma^{-1} \BZ \bw_i$.
    \end{algorithmic}
    \end{algorithm}
\end{figure}

This algorithm serves as the blueprint we aim to translate into a tensor-based setting in this manuscript. Most of its steps have been explained in detail in previous works~\cite{klus_data-driven_2018,klus_data-driven_2020}. We only make the following comments which are important for the following:
\begin{enumerate}
    \item Step~\ref{algline:whitening} serves to compute a reduced basis, called the \emph{whitening transformation}, given as
    \begin{equation}
    \label{eq:whitened_basis}
    \eta(\bx)^\top := \Psi(\bx)^\top \BU \Sigma^{-1}.
    \end{equation}
    By the properties of the SVD, this reduced basis is an empirically orthonormal set of functions. Whitening is equivalent to applying principal component analysis (PCA) on the data matrix $\Psi_\BX$.
    \item Steps~\ref{algline:mean_vector}-\ref{algline:mean_removal} serve to remove the empirical mean from the reduced basis set, and re-orthonormalize afterwards. This eliminates the trivial solution $\kappa_0 = 0$ from the reduced eigenvalue problem.
    \item The calculation of the generator stiffness matrix $\BA$ in line~\ref{algline:stiffness_matrix} is the empirical version of the integration-by-parts formula~\eqref{eq:integration_by_parts}.
\end{enumerate}
Further recent results on generator approximation can be found in Refs.~\cite{devergne_learning_2024,kostic_laplace_2024,devergne_slow_2025}.

\subsection{Reaction Coordinates}
In practice, the basis functions for the VAC are rarely defined as acting on Euclidean molecular coordinates, primarily because molecular systems are translationally and rotationally invariant. Instead, the basis functions are functions of a set of \emph{reaction coordinates} $\bz \in \R^k$, such as internal angles or pairwise distances:

\begin{align*}
    \bz &= \xi(\bx), & \Psi(\bx) &\equiv \Psi(\xi(\bx)) = \Psi(\bz).
\end{align*}
This choice affects the calculation of the generator stiffness matrix $\BA$. Inspecting the integration-by-parts formula~\eqref{eq:integration_by_parts} again, we see that
\begin{equation*}
    \begin{split}
        \BA_{ij} &= \frac{1}{\beta} \int_{\R^d} \nabla \psi_i(\bx)\cdot \Sigma(\bx) \cdot \nabla \psi_j(\bx)\,\diff \mu(\bx) \\
        &= \frac{1}{Z_\beta \beta} \int_{\R^d} \nabla_\bz \psi_i(\bz)\cdot \left[\nabla_\bx \xi(\bx) \Sigma(\bx) \nabla_\bx \xi^\top(\bx)\right] \cdot \nabla_\bz \psi_j(\bx) e^{-\beta V(\bx)}\,\diff \bx \\
        &= \frac{1}{Z_\beta \beta} \int_{\R^k} \nabla_\bz \psi_i(\bz)\cdot \Sigma_{\mathrm{eff}}(\bz) \cdot \nabla_\bz \psi_j(\bz)e^{-\beta F(\bz)} \,\diff \bz, 
    \end{split}
\end{equation*}
where $F$ is the free energy on the reaction coordinate $\bz$. The term $\Sigma_{\mathrm{eff}}(\bz)$ is independent of the basis functions, and can be treated as an effective diffusion field on the reaction coordinates. We see that even if the diffusion field on Euclidean state space is constant, as for the Langevin dynamics, we will obtain an effective position-dependent diffusion in reaction coordinate space, which must be treated efficiently in numerical computations. The empirical estimator for the generator stiffness matrix in line~\ref{algline:stiffness_matrix} must then be adjusted as follows:
\begin{align*}
    \BA_{ij} &= \frac{1}{\beta m}\sum_{l=1}^{m} \nabla_\bz \psi_i(\bz_l) \cdot \Sigma_{\mathrm{loc}}(\bx_l) \cdot \nabla_\bz \psi_j(\bz_l), &
    \Sigma_{\mathrm{loc}}(\bx_l) &= \nabla_\bx \xi(\bx_l) \Sigma(\bx_l) \nabla_\bx \xi^\top(\bx_l),
\end{align*}
where $\bz_l = \xi(\bx_l)$ is just the data projected into the space of reaction coordinates.

Interestingly, this way of approximating the generator in reaction coordinate space is equivalent to analyzing the \emph{coarse-grained} generator implied by the Mori-Zwanzig formalism, so it is an effective way of assessing how suitable the reaction coordinates $\bz$ are in terms of dynamical properties. We refer to Refs.~\cite{zhangEffectiveDynamicsGiven2016a,nuske_spectral_2021,nateghi_kinetically_2025} for a detailed discussion of this topic.

\section{Product Bases and Tensor Representations}
\label{sec:tensor_reps}
Constructing a sufficiently expressive basis set $\Psi$ for high-dimensional systems is non-trivial. In this study, we consider \emph{tensor product} basis sets, which have already been demonstrated to be useful in Refs.~\cite{nuske_tensor-based_2021,lucke_tgedmd_2022,cao_amuset-tica_2025,strand_adaptive_2026}. In this section, we introduce the key concepts and notation, and explain how to implement the whitening transformation on a product basis. This will prepare us to derive a complete tensorized version of Algorithm~\ref{alg:vac} to solve the eigenvalue problem~\eqref{eq:ev_problem_generator}. We begin by describing the construction of the tensor-product basis set.

\subsection{Product Basis: Construction and Notation}
\paragraph{One-dimensional basis functions.}

For each coordinate direction $\bx^k$ (indicated by the super-script), where $k = 1,\ldots,d$ enumerates the components of the state vector $\bx$, let
\[
\Psi^{(k)}(\bx)
=
\big(
\psi^{(k)}_1(\bx^k), \ldots, \psi^{(k)}_n(\bx^k)
\big)^\top
\in \mathbb{R}^n,
\]
be a set of elementary basis functions depending only on the $k$-th coordinate. In this study, we will specifically use a version of \emph{random Fourier features (RFFs)}~\cite{rahimiRandomFeaturesLargescale2007a} as elementary basis sets:
\begin{equation*}
    \psi^{(k)}_{i}(\bx^k) = \frac{1}{\sqrt{2}}\cos(\omega_{i, k} \bx^k + b_{i, k}).
\end{equation*}
The frequencies $\omega_{i,k}$ are randomly sampled from a fixed frequency distribution induced by a kernel function, such as a Gaussian RBF kernel, while the phase shifts $b_{i, k}$ are uniform random numbers in $[0, 2\pi]$. We refer to~\cite{nuske_efficient_2023} for a detailed explanation of this approach, and Refs.~\cite{meanti_estimating_2023,aristoff_fast_2024} for related work.

\paragraph{Multivariate product basis.}

From the elementary basis functions, a multi-variate basis set can be constructed by taking coordinate-wise products:
\begin{equation}
\label{eq:product_basis}
\Psi_{\alpha}(\bx)
=
\prod_{k=1}^d
\psi^{(k)}_{\alpha_k}(\bx^k),
\qquad
\alpha = (\alpha_1,\ldots,\alpha_d).
\end{equation}
Here, the vector $\alpha$ is a \emph{multi-index} with each $\alpha_k \in \{1,\ldots,n\}$. Enumerating all product basis functions yields a vector-valued map $\Psi(\bx) \in \mathbb{R}^{N}$, where the total number of basis functions is $N = n^d$. In line with previous terminology, the \emph{data tensor}
\begin{equation}
\label{eq:data_tensor}
    \Psi_\BX
\in
\mathbb{R}^{N \times m},
\qquad
(\Psi_\BX)_{\alpha, l}
=
\Psi_\alpha(\bx_l),
\end{equation}
collects the evaluations of all product basis functions at all data points.

We note that the partial derivative of a product basis function with respect to $\bx^k$ is
\begin{equation}
\label{eq:product_rule_partial_diff}
\frac{\partial}{\partial \bx^k}
\Psi_{\alpha}(\bx)
=
\left(
\frac{d}{d\bx^k}
\psi^{(k)}_{\alpha_k}(\bx^k)
\right)
\prod_{j\neq k}
\psi^{(j)}_{\alpha_j}(\bx^j).
\end{equation}
We will make use of this structure later to derive low-rank representations for expressions involving the generator.

\subsection{Tensors and Data Representation}
The array $\Psi_\BX$ of product basis evaluations in~\eqref{eq:data_tensor} can be seen as a \emph{tensor} or multi-dimensional array:
\begin{equation*}
    \Psi_\BX \in \R^{n \times n \times \ldots \times n \times m}.
\end{equation*}
The number of independent indices is called \emph{order} of the tensor. We refer to each independent index as a \emph{mode} or \emph{site}, and the dimension of each mode is called \emph{mode size}. Thus, $\Psi_\BX$ is a tensor of order $d + 1$, with the first $d$ mode sizes equal to $n$, and the last one equal to the data size $m$. 

\paragraph{Tensors of Vectors}
The simplest type of tensors are \emph{rank-one} tensors, which are outer products of vectors $\bt^{(k)}\in \R^{n}$:
\[
\BT = \bt^{(1)} \otimes \cdots \otimes \bt^{(d)}.
\]
Every tensor of fixed order can be written as a sum of a finite number of rank-one tensors (\emph{canonical} decomposition). A more flexible and numerically stable representation, however, is provided by the \emph{tensor-train (TT)} format~\cite{oseledets_tensor-train_2011}. A TT tensor of order $d$ is parametrized by $d$ three-dimensional arrays, called \emph{cores}, which are written as abstract matrices~\cite{kazeev_low-rank_2012}:
\begin{equation*}
    \BT^{(k)} = \left\llbracket \begin{matrix}
        \BT^{(k)}_{1,:,1} & \cdots & \BT^{(k)}_{1,:,r_k} \\
\vdots & \ddots & \vdots \\
\BT^{(k)}_{r_{k-1},:,1} & \cdots & \BT^{(k)}_{r_{k-1},:,r_k}
    \end{matrix}\right\rrbracket.
\end{equation*}
Each entry of these core matrices is already a vector of dimension $n$. The row- and column-dimensions $r_k$ of the core matrices are called \emph{TT-ranks} and must satisfy that $r_0 = r_d = 1$. The full tensor is then obtained as a formal matrix product of the cores, with the scalar multiplication replaced by taking outer products between vector entries:
\begin{equation*}
    \BT = \BT^{(1)} \otimes \BT^{(2)} \otimes \cdots \otimes \BT^{(d)}.
\end{equation*}
This is called \emph{matrix-product-state (MPS)} formula~\cite{ostlund_thermodynamic_1995}. Written out entry-wise, the MPS formula is equivalent to:
\begin{equation}
\label{eq:tt_definition}
\BT(\alpha_1,\ldots,\alpha_p)
=
\sum_{l_1=1}^{r_1}\cdots\sum_{l_{p-1}=1}^{r_{p-1}}
\BT^{(1)}_{1,\alpha_1,l_1}
\BT^{(2)}_{l_1,\alpha_2,l_2}
\cdots
\BT^{(p)}_{l_{p-1},\alpha_p,1}.
\end{equation}

\paragraph{Tensors of Matrices} Importantly, tensor formats and operations can be applied to matrices just the same way as to vectors. This will enable the represention of the stiffness matrix $\BA$ using tensor formats. The only difference for tensors of matrices is that the outer product of vectors is replaced by the \emph{Kronecker product} of matrices. To better differentiate matrix-tensors from vectorial tensors of order $d = 2$ (matrices), we refer to them as \emph{tensor operators} (as they correspond to linear operators on tensor spaces).

A TT-operator is parametrized by $d$ four-dimensional cores
\begin{equation*}
    \BS^{(k)} = \left\llbracket \begin{matrix}
        \BS^{(k)}_{1,:,:,1} & \cdots & \BS^{(k)}_{1,:,:,r_k} \\
\vdots & \ddots & \vdots \\
\BS^{(k)}_{r_{k-1},:,:,1} & \cdots & \BS^{(k)}_{r_{k-1},:,:,r_k}
    \end{matrix}\right\rrbracket,
\end{equation*}
where each entry is already a matrix of dimension $n \times n$. The MPS formula remains exactly the same, only using the Kronecker product:
\begin{equation*}
    \BS = \BS^{(1)} \otimes \BS^{(2)} \otimes \cdots \otimes \BS^{(d)}.
\end{equation*}

\subsection{Whitening Transformation in TT Format: Global SVD}
\label{subsec:whitening}

For a product basis~\eqref{eq:product_basis}, the whitening step in line~\ref{algline:whitening} of Algorithm~\ref{alg:vac} can be replaced by a multi-step procedure that only operates on one elementary basis set at a time. The method is called \emph{global SVD}~\cite{klus_tensor-based_2018} and is based on the higher-order SVD for general tensor trains~\cite{oseledets_tensor-train_2011}. A complete description of the method is provided as Algorithm~SI-1 in the appendix. We emphasize the following points:

\begin{enumerate}
    \item The global SVD produces a decomposition $\Psi_\BX \approx \BU \Sigma \BV^\top$, where $\BU \in \R^{n \times n \times \ldots \times n \times r}$ is a TT approximation to the left singular vectors of the data tensor. The TT ranks of $\BU$ can be fixed a priori or controlled adaptively using truncation thresholds. The remaining components $\Sigma \in \R^{r \times r}$ and $\BV \in \R^{m \times r}$ are matrices just as in a standard SVD.
    \item The left-singular tensor $\BU$ is an open-ended TT tensor, meaning that its final core $\BU^{(d)}$ has $r$ columns:
    \begin{equation*}
        \BU^{(d)} = \left\llbracket \begin{matrix}
            \BU^{(d)}_{:, :, 1} & \cdots & \BU^{(d)}_{:, :, r}
        \end{matrix}  \right\rrbracket.
    \end{equation*}
    Each of these columns encodes a proper TT tensor $\BU_j$ which represents a single reduced basis function, in complete analogy to Eq.~\eqref{eq:whitened_basis}, by a TT-structured transformation of the product basis~\cite{nuske_tensor-based_2021}:
    \begin{equation}
    \label{eq:reduced_basis_TT}
    \eta_j(\bx)^\top := \Psi(\bx)^\top \BU_j \sigma_j^{-1} = \Psi(\bx)^\top \left[\BU^{(1)} \otimes \ldots \BU^{(d-1)} \otimes \BU^{(d)}_{:, :, j}\right] \sigma_j^{-1}.
    \end{equation}
    
    \item A crucial observation is that all reduced basis functions share the first $d-1$ cores. This will be exploited in numerical operations on $\BU$ later.
    \item Lines~(\ref{algline:mean_vector}-\ref{algline:mean_removal}) in the VAC Algorithm~\ref{alg:vac} (removal of the mean) can be realized without modification  by acting on the right singular matrix $\BV$.
\end{enumerate}

\section{Tensor-based VAC}
\label{sec:main_algorithm}
In this section, we present our main technical and algorithmic results. The essential remaining step towards a tensorization of Algorithm~\ref{alg:vac} is the assembly of the reduced matrix $\BM$ in lines~(\ref{algline:stiffness_matrix}-\ref{algline:reduced_matrix}). Up to a small-scale post-multiplication, this requires computation of tensor-operator-tensor products of the form
\begin{equation}
\label{eq:auxiliary_reduced_matrix}
    \tilde{\BM} = \BU^\top \BA \BU = \frac{1}{\beta m}\sum_{l=1}^{m} \BU^\top \nabla \Psi(\bx_l) \cdot \Sigma(\bx_l) \cdot \nabla \Psi(\bx_l) \BU.
\end{equation}
Below, we discuss two approaches to evaluating these products using tensor operations. Details are provided in the SI appendix. Algorithm~\ref{alg:tt_vac} summarizes the procedure.

\begin{figure}
    \begin{algorithm}[H]
    \caption{Variational Approximation of Generator Eigenpairs (tensor gEDMD)}\label{alg:tt_vac}
    \begin{algorithmic}[1]
    \Require Elementary basis sets $\Psi^{(k)}$; training data $\bx_1,\ldots, \bx_m$; diffusion $\Sigma(\bx_l)$ for all data points.
    \Ensure Approximate eigenvalues $\kappa_i$ and eigenfunctions $\varphi_i$ solution of~\eqref{eq:ev_problem_generator}
    \algrule
    \State Evaluate elementary basis sets $\Psi^{(k)}(\bx_l)$ and derivatives $\pd{}{\bx^k}\Psi^{(k)}(\bx_l)$ at all data points.
    \State Apply global SVD (Algorithm S1) to data tensor: $\frac{1}{\sqrt{m}}\Psi_\BX
\approx
\BU \Sigma \BV^\top$.
    \State Mean vector of reduced basis: $\bar{\bz} = \frac{1}{m} \BV^\top \mathbf{1}_m \in \mathbb{R}^{r}$.
    \State Mean-free correction to whitening transform: diagonalize $\mathrm{Id}_r  - \bar{\bz}\bar{\bz}^{\top} = \BW \,\mathrm{diag}(\lambda)\, \BW^{\top}$ and set $\BZ = \BW \mathrm{diag}(\lambda^{-1/2})$.
    \State Preliminary reduced matrix using either Algorithm~S3 or~S4: $\tilde{\BM}_{ij} = \BU_i^\top
\BA \BU_j$.
    \State Final reduced generator matrix: $\BM = \BZ^\top \Sigma^{-1} \tilde{\BM} \Sigma^{-1} \BZ \in \R^{r \times r}$.
    \State Small-scale eigenvalue problem: $\BM \bw_i = \kappa_i \bw_i$.
    \State Evaluate eigenfunctions: $\varphi_i(\bx) = \Psi(\bx)^\top \BU \Sigma^{-1} \BZ \bw_i$.
    \end{algorithmic}
    \end{algorithm}
\end{figure}

\subsection{Operator-Tensor Products}
\label{subsec:operator_tensor_products}
The first option we consider is to represent the operator $\BA$ in TT format and, since $\BU$ is also a TT tensor, evaluate~\eqref{eq:auxiliary_reduced_matrix} using established algorithms for operator-tensor-products in TT format. We show in SI Appendix 2 that $\BA$ has an exact TT-operator representation with cores $\BA^{(k)}, 1 \leq k \leq d$, with ranks scaling as $(2k + 2)m$ (for constant diffusion, this TT-rank reduces to $2m$ uniformly). The analytical expressions to form these cores are highly sparse and only require access to the elementary basis functions, their gradients, and the diffusion field $\Sigma$. In SI Appendix 3.1, we then explain how sparsity and the structure of $\BU$ can be exploited to evaluate~\eqref{eq:auxiliary_reduced_matrix} with a minimal number of operations.

\subsection{Gradient Contraction}
\label{subsec:gradient_contractions}
As an alternative, we observe from the second expression in~\eqref{eq:auxiliary_reduced_matrix} that the main computational bottleneck is the computation of vectors
\begin{equation*}
    \bw^{(i)}_l = \BU^\top \pd{}{\bx^i}\Psi(\bx_l) \in \R^r,
\end{equation*}
for every sample point $\bx_l$ and state dimension $1 \leq i \leq d$. This evaluation boils down to a contraction of the tensor train $\BU$ with a rank-one tensor $\pd{}{\bx^i}\Psi(\bx_l)$. Because of the product rule~\eqref{eq:product_rule_partial_diff}, a streamlined evaluation of all vectors $\bw^{(i)}_l$ can be designed efficiently. The details for this procedure are provided in SI appendix 3.2. In the numerical examples, we test both approaches presented here and compare their performance.

\subsection{Updating the Reduced Matrix}
Suppose we have already computed the global SVD with components $\BU, \Sigma, \BV$, as well as the reduced matrix $\tilde{\BM}$ in Eq.~\eqref{eq:auxiliary_reduced_matrix} for a product basis on $d$ state variables $\bx^1, \ldots, \bx^d$. We can add another state variable, say $\bx^{d+1}$, along with its elementary basis set $\Psi^{(d+1)}$, to the full product basis, without re-running the entire computation from scratch. This can be useful to test if the variable $\bx^{d+1}$ is relevant to capture slow conformational changes. The required steps are summarized in Algorithm~\ref{alg:update_m}. It essentially takes one additional matrix SVD to update the global SVD (Algorithm~S1), followed by an extension of the gradient contractions $\bw_l^{(i)}$ to the enlarged basis, and a single reassembly of $\tilde{\BM}$.

Since the SVD update in line~\ref{algline:svd_update} leaves the existing cores $\BU^{(1)},\ldots,\BU^{(d)}$ unchanged and introduces only the new core $\BU^{(d+1)}$, the gradient contractions $\bw_l^{(i)} = \BU^\top \pd{}{\bx^i}\Psi(\bx_l)$ for $1 \leq i \leq d$ only need to be contracted against the new core $\BU^{(d+1)}$. The projection $\bw_l^{(d+1)}$ associated with the new dimension, by contrast, has no previous counterpart. Obtaining these vectors requires loading or re-computing the base carries $\bq_l$ in Algorithm~S4, and contracting them against the new basis set derivatives $\td{}{\bx^{d+1}}\Psi^{(d+1)}$. Once all $d+1$ projections $\bw_l^{(i)}$ are available, $\tilde{\BM}$ is re-computed using the extended diffusion factors $\Sigma(\bx_l)\in\R^{(d+1)\times(d+1)}$ exactly as in Eq.~(S7) (line~20 of Algorithm~S4).

\begin{figure}
    \begin{algorithm}[H]
    \caption{Update Reduced Matrix}\label{alg:update_m}
    \begin{algorithmic}[1]
    \Require Elementary basis set $\Psi^{(d+1)}$; training data $\bx_1,\ldots, \bx_m$; extended diffusion factors $\Sigma(\bx_l) \in \R^{(d+1) \times (d+1)}$ for all data points; existing cores $\BU^{(1)},\ldots,\BU^{(d)}$ and existing projections $\bw_l^{(i)}$, $i=1,\ldots,d$, $l=1,\ldots,m$, from the previous assembly of $\tilde{\BM}$.
    \Ensure Updated reduced matrix $\tilde{\BM}$ for the product basis including $\Psi^{(d+1)}$.
    \algrule
    \State Perform one step of the global SVD Algorithm~S1 to compute core $\BU^{(d+1)}$ and update $\Sigma,\, \BV$; let $r'$ denote the resulting (possibly updated) reduced rank. \label{algline:svd_update}
    \For{$l = 1,\ldots,m$}
        \State Extend existing projections: $\bw_l^{(i)} \gets \big((\bw_l^{(i)})^\top \BU^{(d+1)}\big)^\top \Psi^{(d+1)}(\bx_l^{d+1})$, for $i=1,\ldots,d$  \label{algline:extend_old}
        \State Re-compute or load base carry $\bq_l$.\label{algline:base_recompute}
        \State Compute new projection: $\bw_l^{(d+1)} \gets \big(\bq_l^\top \BU^{(d+1)}\big)^\top \td{}{\bx^{d+1}}\Psi^{(d+1)}(\bx_l^{d+1})$ \Comment{differentiate at new site, as in line~9 of Alg.~S4} \label{algline:new_projection}
    \EndFor
    \State Collect the updated $\bw_l^{(i)}$, $i=1,\ldots,d+1$, $l=1,\ldots,m$, into $\BW \in \R^{(m(d+1))\times r'}$ \label{algline:store_update}
    \State Reassemble $\tilde{\BM} \gets -\dfrac{1}{2m}\, \BW^\top(\Sigma\BW)$, with $\Sigma(\bx_l)\in\R^{(d+1)\times(d+1)}$ \label{algline:reassemble}
    \end{algorithmic}
    \end{algorithm}
\end{figure}

\FloatBarrier

\subsection{TICA Projection}

In the examples presented below, we will choose the state variables $\bx^k$ as TICA coordinates (TICs)~\cite{molgedey_separation_1994,perez-hernandez_identification_2013}. This is in line with previous studies, which have shown that TICs often coarsely capture the slow dynamics of molecular systems, but timescale estimates and PCCA decompositions can be improved significantly by an additional VAC step~\cite{mardt_vampnets_2018,cao_amuset-tica_2025,sprink_accurate_2026}. As TICs are usually derived from non-linear features like distances or angles, it follows that the diffusion field $\Sigma_{\mathrm{loc}}$ is state-dependent. The diffusion tensors can be pre-computed once from the MD simulation data before applying the tensor-based method.

TICA coordinates also come with a natural ordering of the state variables, which is a challenge for tensor-based methods when working on general coordinates. Moreover, application of the updating scheme provides an efficient test of how many TICs are required to capture the main conformational dynamics of the system at hand.

\section{Results}\label{sec:Applications}

\subsection{Systems and Data Generation}
We study three examples: a minimal three-dimensional model potential, as well as molecular dynamics simulation data of the fast-folding proteins Chignolin and NTL9.

The toy model is Langevin dynamics in a three-dimensional energy landscape. The potential is derived from the two-dimensional lemon slice potential
\begin{equation*}
    V_{\mathrm{LS}}(\bx^1, \bx^2) = 10(r - 1)^2 + \cos(4\varphi),
\end{equation*}
where $r, \varphi$ are polar coordinates. This 2d-system is augmented by a simple harmonic potential along the third state variable
\begin{equation*}
    V_{\mathrm{h}} = \frac{1}{2}\alpha (\bx^3)^2,
\end{equation*}
and the total potential is $V = V_{\mathrm{LS}} + V_{\mathrm{h}}$. The third dimension does not add any dynamical content to this example, we only add the third dimension to turn this into an example where tensor-based methods are relevant, while we can still easily compute full matrix-based reference quantities to compare to.

Simulations of the fast-folding mini-protein Chignolin have served as a benchmark in many studies on molecular kinetics in recent years. The simulation data we use consists of 20 independent $5 \mathrm{\mu s}$ long simulations provided by the Theoretical and Computational Biophysics Group at Freie Universität Berlin, see Ref.~\cite{charron_navigating_2025} for the simulation setup. It is well-known that performing a TICA analysis on all 45 pairwise carbon-alpha distances provides a two-dimensional TICA space which resolves Chignolin's three major metastable states: folded, unfolded, and misfolded.

Simulations of the 39-residue protein NTL9 were provided by D.E. Shaw Research as described in Ref.~\cite{lindorff-larsen_how_2011}. The simulation data consist of four independent simulations with a combined length of $1.1\,\mathrm{ms}$. We apply TICA on the set of all pairwise carbon-alpha distances. The protein's major metastable states correspond to its folded, unfolded, an intermediate, and an off-pathway misfolded state.

The two molecular simulation data sets were generated using thermostatted MD in position and momentum space. However, it is well-known that their projection along low-dimensional reaction coordinates can be described by a reversible diffusive dynamics on position space only, after a suitable re-scaling of time, see Refs~\cite{nuske_spectral_2021,nateghi_kinetically_2025} for a detailed discussion. Therefore, we apply reversible tensor gEDMD as described in Section~\ref{sec:tensor_reps}. The resulting implied timescales need to be understood in re-scaled time units.

\begin{figure}[H]
    \centering
    \includegraphics[width=\textwidth]{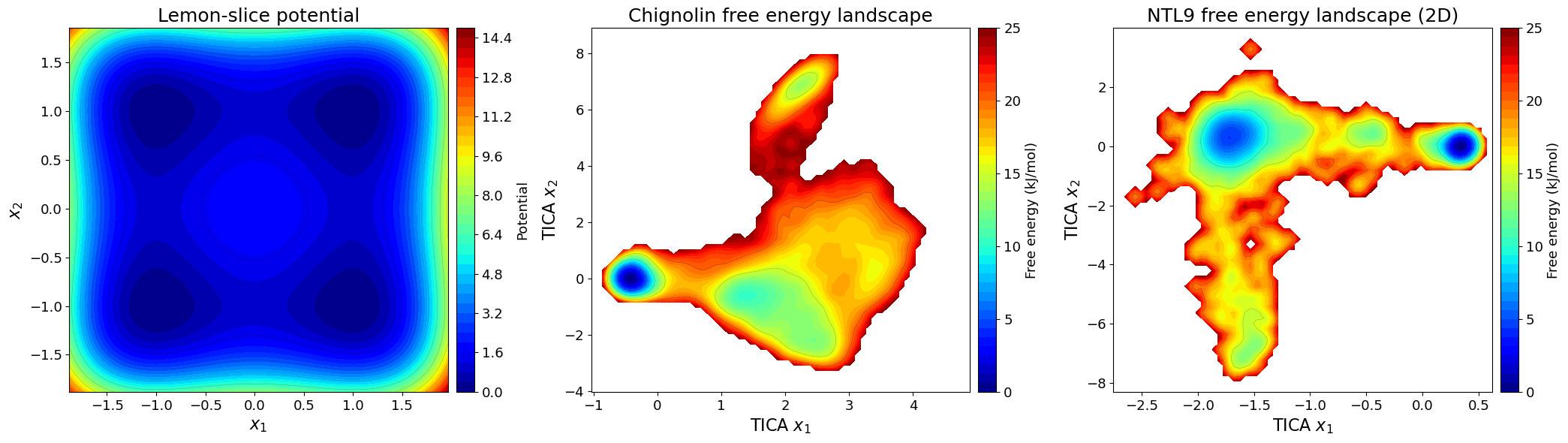}
    \caption{
    Comparison of free energy landscapes for the synthetic lemon slice potential and two protein systems. 
    \textbf{Left:} The analytical two-dimensional lemon-slice potential used as a benchmark, shown as a contour plot in the coordinates $(\bx^1,\bx^2)$. 
    \textbf{Middle:} Two-dimensional free-energy landscape of Chignolin projected onto the first two TICA coordinates $(\mathrm{TICA}\;\bx^1,\mathrm{TICA}\;\bx^2)$. 
    \textbf{Right:} Corresponding two-dimensional free-energy landscape of NTL9 in the first two TICA coordinates. 
    In all panels, lower free-energy regions (blue) correspond to highly populated metastable states, while higher free-energy regions (red) indicate less frequently sampled conformations.
    }
    \label{fig:intro_free_energy_landscapes}
\end{figure}

\subsection{Computational Parameters}
The main computational parameters of the tensor gEDMD method are the number of data points $m$, the choice of kernel function and its hyper-parameters, as well as the truncation parameter for the global SVD algorithm.
We employ a Gaussian radial basis function kernel with a single hyper-parameter, its bandwidth $\sigma_\omega$. Table \ref{tab:comp_params} shows the data sizes we used for each system, as well as the range of bandwidths and truncation parameters we tested.

\begin{table}[htbp]
\centering
\caption{Computational parameters tested for the tensor gEDMD method
across the three benchmark systems.
\label{tab:comp_params}}

\resizebox{\textwidth}{!}{%
\begin{tabular}{@{}l l l l l@{}}
\toprule
System & Data size $m$ & Bandwidth $\sigma_\omega$ & TT rank & SVD truncation threshold \\
\midrule
3D Lemon Slice Potential
   & $7000$
  & $4.0$
  & $400$
  &  $10^{-9}$ \\[2pt]

Chignolin
  & $6000$
  & $25.0$
    & no cap
   
  & $10^{-12}$ \\[2pt]
NTL9
  & $6485$
  & $25.0$
  & no cap
  & $10^{-12}$ \\
\bottomrule
\end{tabular}%
}

\end{table}
\FloatBarrier

\subsection{Model Potential}
We first validate our method on the two-dimensional lemon slice potential, which serves as a controlled benchmark with a known metastable structure. The potential consists of four energetically favorable wells separated by barriers. Since the dynamics evolve in a low-dimensional state space, a matrix representation of the generator can be constructed explicitly for the product basis, which serves as reference model to compare to.

\begin{figure}[H]
    \centering
    \includegraphics[width=\textwidth]{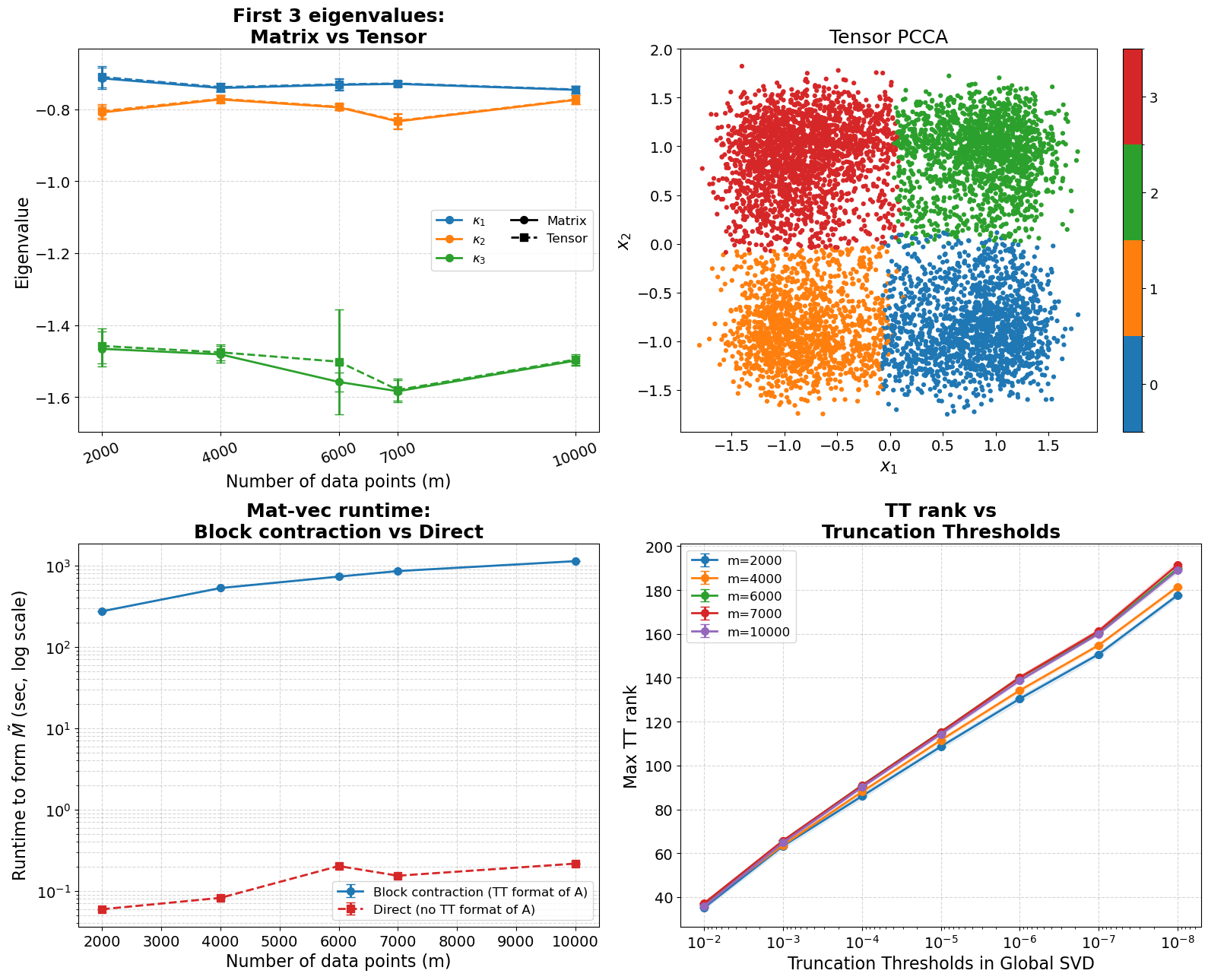}
    \caption{
    Numerical validation of the tensor-based generator approximation for the two-dimensional lemon slice benchmark.
    (\textbf{Upper left}) Comparison of the first three dominant generator eigenvalues obtained using the direct matrix formulation and the tensor-based formulation as functions of the number of sampled data points $m$. Results are averaged over ten random subsamples of a single long Langevin trajectory, with error bars representing one standard deviation.
    (\textbf{Upper right}) Four-state PCCA decomposition computed from the dominant eigenspace of the tensor-based generator approximation, accurately recovering the metastable regions associated with the four potential wells.
    (\textbf{Lower left}) Wall-clock time required to assemble the reduced generator $\tilde{M}$ as a function of the number of sampled data points $m$, comparing the block-contraction scheme (Section~\ref{subsec:operator_tensor_products}) against the direct scheme (Section~\ref{subsec:gradient_contractions}).
    (\textbf{Lower right}) Maximum tensor-train (TT) rank as a function of the global SVD truncation threshold for different sample sizes.
    }
    \label{fig:toy_results}
\end{figure}

Figure~\ref{fig:toy_results} summarizes the numerical results. In the upper-left panel, we compare the first three non-trivial generator eigenvalues obtained from the tensor-based and direct matrix-based formulations for increasing numbers of sampled data points. For each sample size, the computation is repeated on ten random subsamples extracted from a single long Langevin trajectory. The tensor-based eigenvalues are close to indistinguishable from those obtained using the matrix formulation across all sample sizes, with the observed differences remaining well within the statistical uncertainty indicated by the error bars. Furthermore, the dominant eigenvalues remain stable as the amount of available data increases, indicating that the tensor representation does not compromise the accuracy or convergence of the generator approximation. Based on the dominant eigenfunctions of the tensor-based generator approximation, a four-state PCCA decomposition is computed. As illustrated in the upper-right panel, the recovered partition accurately identifies the four metastable basins together with their separating transition regions.

The lower-left panel compares the wall-clock time required to assemble the reduced generator $\tilde{M}$ using two different tensor-train evaluation strategies: the block-contraction scheme from Section~\ref{subsec:operator_tensor_products}, which forms $\tilde{M}$ via repeated tensor-train matrix-vector products, and the direct scheme based on gradient contractions from Section~\ref{subsec:gradient_contractions}. Across the full range of sample sizes considered, the direct scheme outperforms the block-contraction scheme by roughly three to four orders of magnitude. Both schemes yield numerically identical reduced operators, but the cost of assembling $\tilde{M}$ is significantly different.

The computational complexity of the tensor representation is further illustrated in the lower-right panel, which reports the maximum TT rank as a function of the truncation threshold employed during the global SVD. As expected, stricter truncation thresholds retain more singular vectors and therefore result in larger TT ranks. More importantly though, the dependence of the TT rank on the number of sampled data points is very weak, suggesting that the complexity of the tensor representation is primarily governed by the intrinsic low-rank structure of the problem rather than by the amount of available data. Overall, these results demonstrate that the tensor-based formulation accurately reproduces both the spectral properties and the metastable decomposition of the model system, and that the reduced generator can be assembled efficiently.

\subsection{Chignolin MD Simulation Data}

Next, we apply the tensor gEDMD framework on molecular dynamics simulation data of the mini-protein Chignolin, which has been used as a benchmark for numerous kinetic modeling approaches in recent years. As its main conformational states (folded, unfolded, and misfolded) can already be captured by just two TICA coordinates, the main point of this example is to show that tensor gEDMD reliably identifies these states and their transition timescales for a wider range of TICA dimensions and under variations of hyper-parameters. We also find that the bandwidth parameter for the random Fourier basis set needs to be adjusted as the number of TICA dimensions is increased.

\begin{figure}[H]
    \centering
    \includegraphics[width=\textwidth]{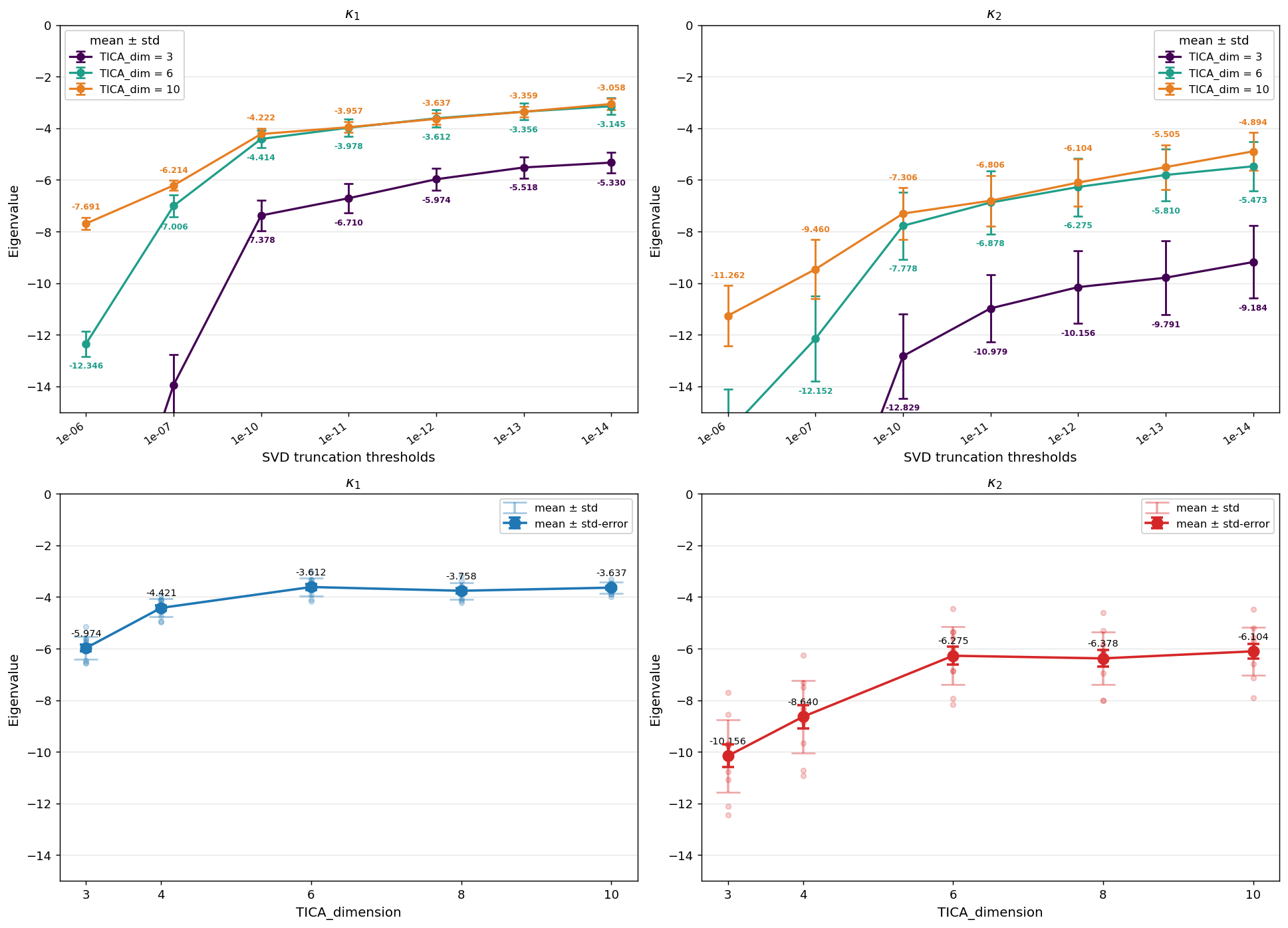}
    \caption{
    Sensitivity of the tensor-based generator approximation to the global SVD truncation threshold and the retained TICA dimension for the Chignolin data set ($m=6000$, ten independent subsamples).
    (\textbf{Upper left}) Dependence of the dominant generator eigenvalue $\kappa_1$ on the global SVD truncation threshold for TICA dimensions $d=3$, $6$, and $10$.
    (\textbf{Upper right}) Corresponding results for $\kappa_2$.
    The annotated values indicate the mean eigenvalue estimates, while the error bars represent one standard deviation across subsamples.
    (\textbf{Lower left}) Dependence of $\kappa_1$ on the number of retained TICA coordinates for a fixed truncation threshold of $10^{-12}$.
    (\textbf{Lower right}) Corresponding dependence of $\kappa_2$.
    Large markers denote the sample mean, thick error bars indicate one standard deviation, thin error bars represent the standard error, and the transparent markers correspond to the individual subsample estimates.
    }
    \label{fig:protein_svd_tica}
\end{figure}

\begin{figure}[H]
    \centering
    \includegraphics[width=\textwidth]{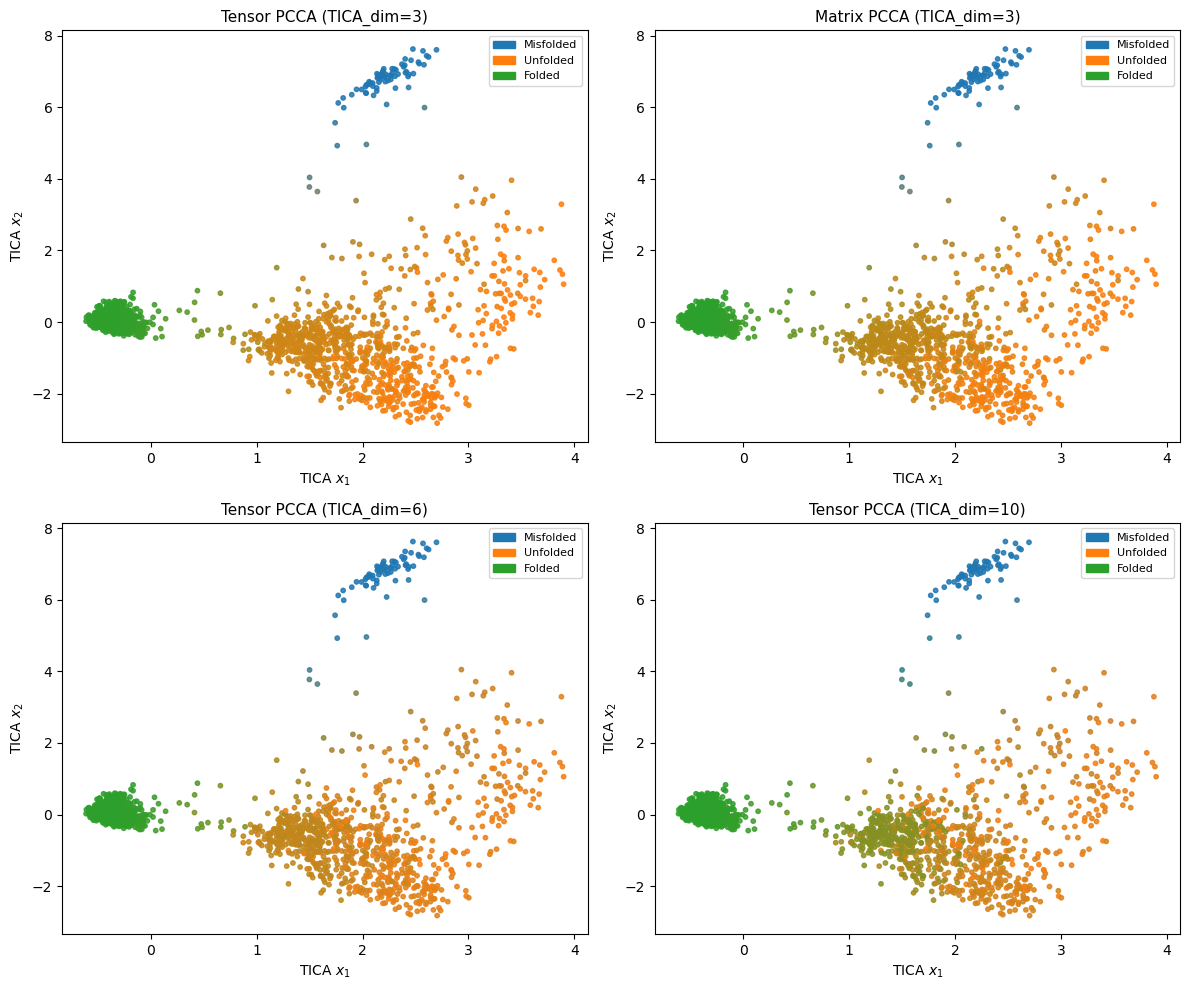}

\caption{PCCA soft membership decomposition of the Chignolin data projected 
onto the first two TICA coordinates. The top panels compare the 
tensor train and matrix-based formulations for $d = 3$ TICA coordinates. The lower panels show 
tensor train PCCA decompositions for $d = 6$ and $d = 10$ TICA coordinates. Pure colors indicate 
configurations assigned predominantly to a single metastable state, while 
mixed colors indicate transition regions.}
    \label{fig:protein_pcca_tica}
\end{figure}

In Figure~\ref{fig:protein_svd_tica}, we show the dependence of the two dominant non-trivial generator eigenvalues, $\kappa_1$ and $\kappa_2$, on the global SVD truncation threshold for multiple TICA dimensions $d=3$, $6$, and $10$, while keeping the bandwidth parameter fixed at $\sigma_\omega = 25$. We can identify a regime of threshold parameters $\epsilon \leq 10^{-10}$ for which both eigenvalue estimates converge, and the leading rate $\kappa_1$ is approximately a factor $\frac{1}{2}$ smaller than $\kappa_2$, in line with previous results. We see that as a function of the TICA dimension, eigenvalue estimates stabilize if $d \geq 6$, while the results are sub-optimal for smaller $d$. This observation is plausible as a larger bandwidth is required to accurately distinguish states embedded in higher-dimensional feature spaces. In Supplementary Figure S1, we show that a smaller bandwidth leads to accurate results in lower TICA dimension $d$. Figure~\ref{fig:protein_pcca_tica} shows membership decompositions computed by the PCCA algorithm, visualized along the first two TICA coordinates. We confirm once again that for TICA dimension $d = 3$, the tensor-train and matrix-based formulations produce nearly identical decompositions. Tensor gEDMD results obtained for larger TICA dimensions are likewise in agreement with the expected PCCA decomposition, although the unfolded state appears somewhat blurred out by these models.

%

\paragraph{Incremental update versus re-computation from scratch.}

\begin{figure}[H]
    \centering
    \includegraphics[width=\textwidth]{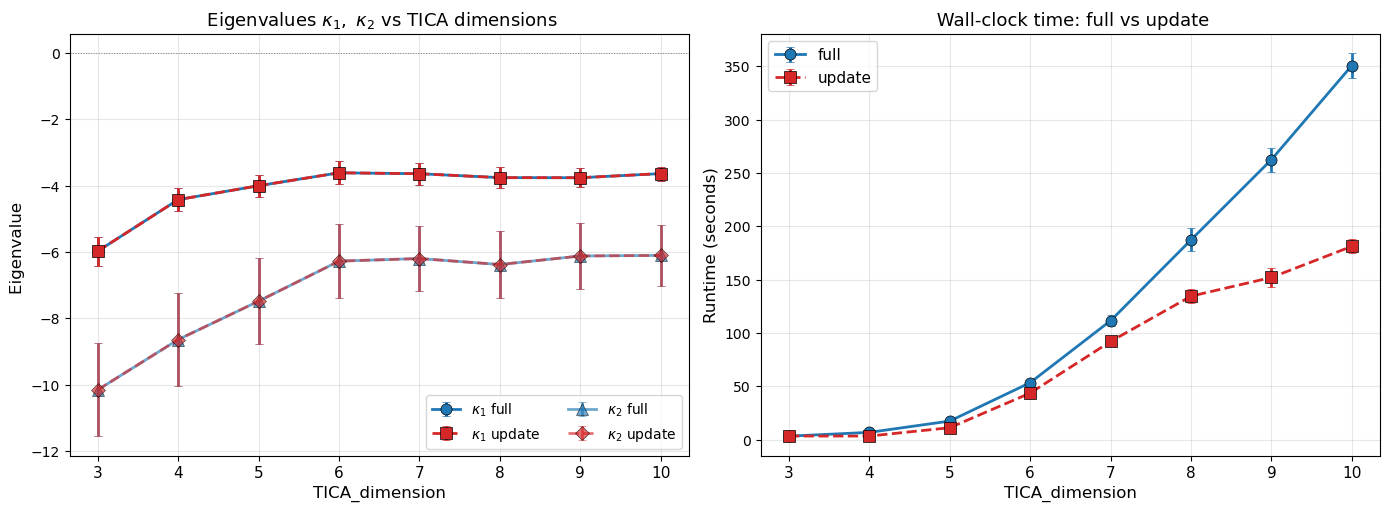}
    \caption{
    Comparison of the incremental tensor-train update with full recomputation for the Chignolin data set.
    (\textbf{Left}) The two dominant non-trivial generator eigenvalues, $\kappa_1$ and $\kappa_2$, computed using the full tensor construction and the incremental update as functions of the retained TICA dimension. Results are averaged over ten independent subsamples, with error bars representing one standard deviation. The two approaches produce nearly identical eigenvalue estimates across all considered dimensions.
    (\textbf{Right}) Wall-clock runtime of the full re-computation and the incremental update for the same experimental settings. The incremental update consistently outperforms full re-computation, and the run-time advantage becomes more pronounced as the TICA dimension increases due to the reuse of the previously computed tensor-train representation.
    }
    \label{fig:runtime_recompress}
\end{figure}

We also illustrate the effect of the incremental updating method provided in Algorithm~\ref{alg:update_m} for the Chignolin example. First, in the left panel of Figure~\ref{fig:runtime_recompress}, we confirm that both of the leading non-trivial generator eigenvalues, $\kappa_1$ and $\kappa_2$, are re-produced correctly if computed sequentially for each dimension by applying the updating algorithm to the previous result. The computational advantage incurred by the incremental update is illustrated in the right panel of Figure~\ref{fig:runtime_recompress}, which compares the wall-clock runtime of the updating method versus re-computing the tensor gEDMD model from scratch. For each TICA dimension, the same data set, kernel parameters, and SVD truncation settings are used as in the eigenvalue comparison. The run-time advantage is not significant for small $d \leq 6$, but becomes much more pronounced for larger TICA dimension $d \geq 8$.

\subsection{NTL9 MD Simulation Data}

As a third example, we validate the tensor gEDMD framework on simulation data of the NTL9 protein. NTL9 is a more complex system which exhibits multiple relevant conformational states which cannot all be resolved by a simple two-dimensional TICA projection. 

\begin{figure}[H]
    \centering
    \includegraphics[width=.8\textwidth]{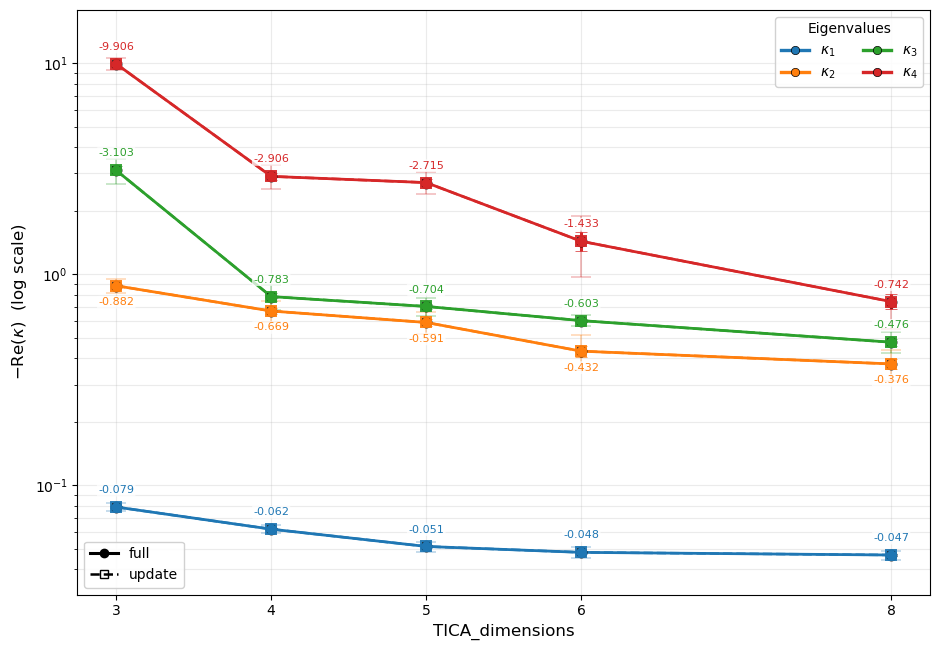}
    \caption{
    Dependence of the first four dominant generator eigenvalues on the retained TICA dimension for the NTL9 data set ($m=6485$, $\sigma_\omega=25$, global SVD truncation threshold $10^{-12}$). Results are averaged over ten independent subsamples, with error bars representing one standard deviation. Solid lines with circular markers correspond to a complete recomputation of the tensor-based generator, while dashed lines with square markers denote the incremental tensor-train update. Across all considered TICA dimensions, the incremental update reproduces the eigenvalues obtained from full recomputation with virtually identical accuracy, indicating that extending the tensor representation incrementally preserves the dominant spectral information of the generator.
    }
    \label{fig:ntl9_tica_dim_eigenvalues_zoom}
\end{figure}

\begin{figure}[H]
\centering
\includegraphics[width=\textwidth]{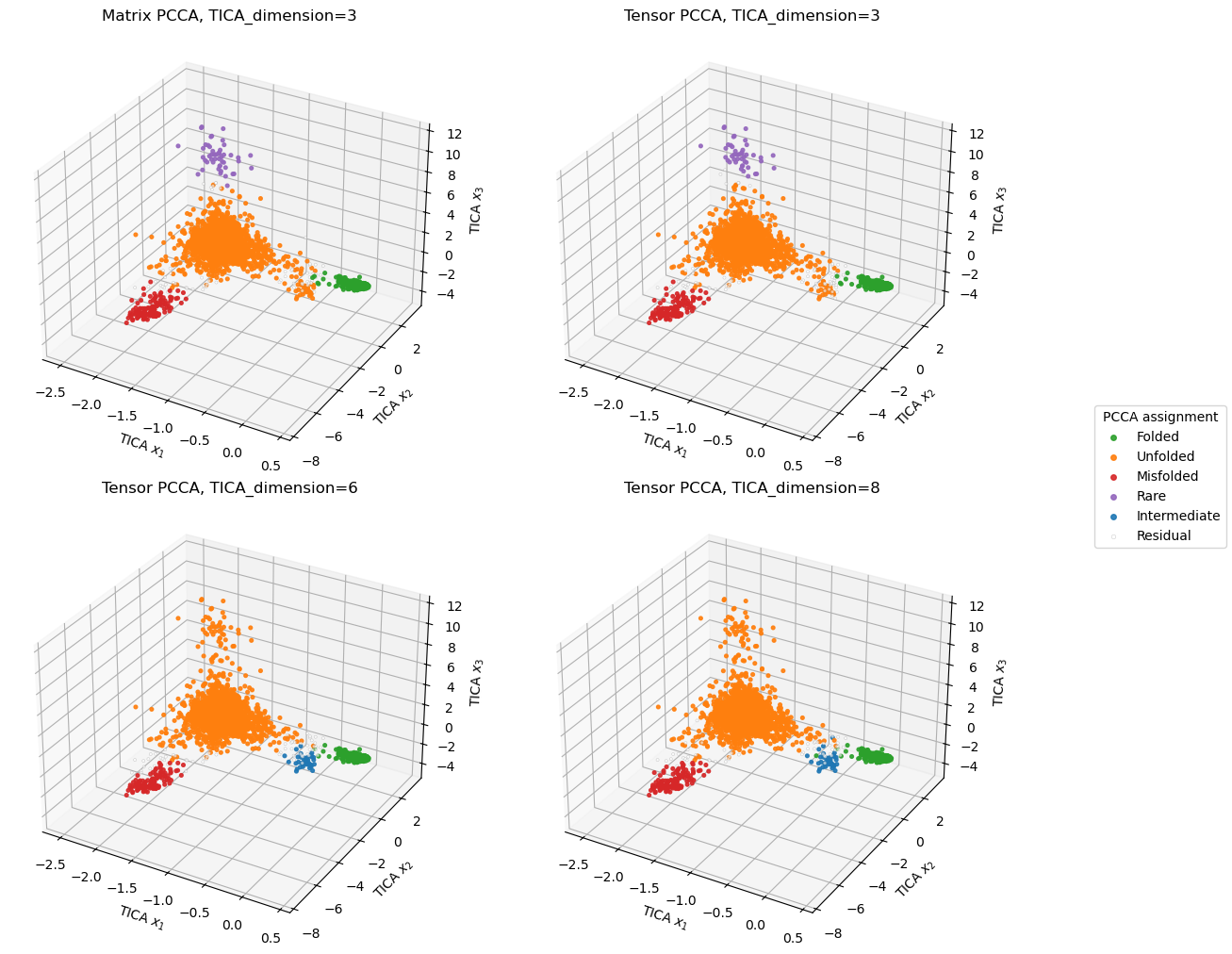}
\caption{Aligned PCCA state assignments for the NTL9 data projected onto the first three TICA coordinates, $\bx^1$, $\bx^2$, and $\bx^3$. The top panels compare the matrix-based and tensor-train formulations for $d=3$, with state labels aligned so that matching colors correspond to comparable metastable regions. The lower panels show tensor-train PCCA decompositions for $d=6$ and $d=8$. Colors indicate the assigned metastable states, while residual points indicate configurations that are not assigned confidently to any metastable state.}
\label{fig:ntl9_pcca_tica_aligned}
\end{figure}

Figure~\ref{fig:ntl9_tica_dim_eigenvalues_zoom} illustrates the dependence of the first four non-trivial generator eigenvalues, $\kappa_1,\ldots,\kappa_4$, on the number of retained TICA coordinates (we also show estimates obtained by the incremental updating strategy, which are virtually the same). The first two eigenmodes are stably recovered for all TICA dimensions $d \geq 3$, with rates around $\kappa_1 \approx 0.05$ and $\kappa_2 \approx 0.5$, respectively. However, we observe that after adding a fourth TICA dimension, an additional mode $\kappa_3$ of about the same magnitude as $\kappa_2$ is discovered, and the remaining spectrum shifts downward accordingly. The leading three modes then remain stable through TICA dimension $d = 8$. In Figure~\ref{fig:ntl9_pcca_tica_aligned}, we show the PCCA state assignments obtained from tensor gEDMD models at different TICA dimensions, projected onto the first three TICA coordinates, $\bx^1$, $\bx^2$, and $\bx^3$. Comparison of these results shows that all models share three conformational states (green, red and orange in Fig.~\ref{fig:ntl9_pcca_tica_aligned}) corresponding to the folded, unfolded and misfolded state of the protein. However, gEDMD models at $d = 3$ identify the next slowest transition as aligned with the third TICA coordinate. Upon addition of the fourth TICA coordinate, a transition to a different state is identified which is in fact aligned with the fourth TICA coordinate. Comparison of the molecular geometry shows that this new state is indeed the intermediate state identified previously~\cite{sprink_accurate_2026}. These observations show that linear TICA is not sufficient to correctly recover all higher modes in the correct order, while the non-linear tensor gEDMD model can identify the correct order of state transitions.

\FloatBarrier

\section{Conclusions and Outlook}\label{sec:Conclusion and Outlook}

In this work, we have developed tensor gEDMD, a variational approximation method to compute metastable states and transition timescales of reversible molecular dynamics simulations employing a product basis. We have derived analytical expressions of the generator stiffness matrix in tensor train format, incorporating state-dependent diffusion fields which often occur when working in reduced coordinates. We have developed efficient algorithms to solve the resulting variational problem using tensor operations, only requiring access to elementary basis functions and their gradients, as well as to the diffusion field, which can be computed as a pre-processing step. We have also described how the tensor gEDMD model can be updated efficiently in order to incorporate additional descriptors.

The numerical results show that tensor gEDMD recovers metastable states and transition timescales for examples including a model potential and two fast-folding protein simulations. The results are stable across different data sizes and truncation parameters. For the Chignolin example, we have observed that the Gaussian kernel bandwidth needs to be chosen carefully depending on the number of input TICA coordinates. For NTL9, we have seen that the non-linear tensor gEDMD model correctly identifies the protein's major metastable states for sufficiently large TICA dimension, improving on the use of the standard linear TICA model.

Several directions remain for future work. First, we will seek to apply the method for larger-scale systems than fast-folding proteins, which exhibit a more complex hierarchy of multiple meta-stable states and transition pathways. Moreover, application to systems driven by underdamped Langevin dynamics can also be studied, building on results from~\cite{nateghi_consistent_2026}. Finally, we will study ways of adaptively tuning key parameters, such as the kernel bandwidth and SVD cutoff, to improve the efficiency and accuracy of the method further.

\section{Declarations}
The authors have no conflicts of interest to disclose.

\paragraph{Data Availability}
Data and codes to reproduce the results presented in this study are available from the Github repository \url{https://github.com/fnueske/tensor_gedmd} and the data repository \url{10.5281/zenodo.22080161}.


\paragraph{Use of Artificial Intelligence} We used ChatGPT (OpenAI) for language and grammar editing, to improve clarity of the manuscript, and to clarify conceptual questions and aid our understanding of topics encountered during the study. Claude (Anthropic) was used to assist with the organization and structuring of code.

\paragraph{Funding and Acknowledgemts} M. Verma is funded by Deutsche Forschungsgemeinschaft (DFG, German Research Foundation) - 314838170, GRK 2297 MathCoRe.
In addition, we are grateful to D.E. Shaw research for providing access to the folding simulations of NTL9.

\bibliography{references}

@article{klus_data-driven_2020,
    title = {Data-driven approximation of the {Koopman} generator: {Model} reduction, system identification, and control},
    volume = {406},
    doi = {10.1016/j.physd.2020.132416},
    journal = {Physica D: Nonlinear Phenomena},
    author = {Klus, Stefan and Nüske, Feliks and Peitz, Sebastian and Niemann, Jan Hendrik and Clementi, Cecilia and Schütte, Christof},
    year = {2020},
}

@article{klus_data-driven_2018,
    title = {Data-{Driven} {Model} {Reduction} and {Transfer} {Operator} {Approximation}},
    volume = {28},
    doi = {10.1007/s00332-017-9437-7},
    number = {3},
    journal = {Journal of Nonlinear Science},
    author = {Klus, Stefan and Nüske, Feliks and Koltai, Péter and Wu, Hao and Kevrekidis, Ioannis and Schütte, Christof and Noé, Frank},
    year = {2018},
    pages = {985--1010},
}

@article{nuske_tensor-based_2021,
    title = {Tensor-based computation of metastable and coherent sets},
    volume = {427},
    doi = {10.1016/j.physd.2021.133018},
    journal = {Physica D: Nonlinear Phenomena},
    author = {Nüske, Feliks and Gelß, Patrick and Klus, Stefan and Clementi, Cecilia},
    year = {2021},
}

@article{cao_amuset-tica_2025,
    title = {{AMUSET}-{TICA}: {A} {Tensor}-{Based} {Approach} for {Identifying} {Slow} {Collective} {Variables} in {Biomolecular} {Dynamics}},
    volume = {21},
    copyright = {https://doi.org/10.15223/policy-029},
    issn = {1549-9618, 1549-9626},
    shorttitle = {{AMUSET}-{TICA}},
    url = {https://pubs.acs.org/doi/10.1021/acs.jctc.5c00076},
    doi = {10.1021/acs.jctc.5c00076},
    language = {en},
    number = {9},
    urldate = {2025-06-18},
    journal = {Journal of Chemical Theory and Computation},
    author = {Cao, Siqin and Nüske, Feliks and Liu, Bojun and Soley, Micheline B. and Huang, Xuhui},
    month = may,
    year = {2025},
    pages = {4855--4866},
}

@article{oseledets_tensor-train_2011,
    title = {Tensor-{Train} {Decomposition}},
    volume = {33},
    issn = {1064-8275, 1095-7197},
    url = {http://epubs.siam.org/doi/10.1137/090752286},
    doi = {10.1137/090752286},
    language = {en},
    number = {5},
    urldate = {2026-03-23},
    journal = {SIAM Journal on Scientific Computing},
    author = {Oseledets, I. V.},
    month = jan,
    year = {2011},
    pages = {2295--2317},
}

@article{klus_tensor-based_2018,
    title = {Tensor-based dynamic mode decomposition},
    volume = {31},
    issn = {0951-7715, 1361-6544},
    url = {https://iopscience.iop.org/article/10.1088/1361-6544/aabc8f},
    doi = {10.1088/1361-6544/aabc8f},
    number = {7},
    urldate = {2026-03-23},
    journal = {Nonlinearity},
    author = {Klus, Stefan and Gelß, Patrick and Peitz, Sebastian and Schütte, Christof},
    month = jul,
    year = {2018},
    pages = {3359--3380},
}

@article{lucke_tgedmd_2022,
	title = {{tgEDMD}: {Approximation} of the {Kolmogorov} {Operator} in {Tensor} {Train} {Format}},
	volume = {32},
	doi = {10.1007/s00332-022-09801-0},
	number = {4},
	journal = {Journal of Nonlinear Science},
	author = {Lücke, Marvin and Nüske, Feliks},
	year = {2022},
}

@article{nuske_variational_2014,
    title = {Variational approach to molecular kinetics},
    volume = {10},
    doi = {10.1021/ct4009156},
    number = {4},
    journal = {Journal of Chemical Theory and Computation},
    author = {Nüske, Feliks and Keller, Bettina G. and Pérez-Hernández, Guillermo and Mey, Antonia S. J. S. and Noé, Frank},
    year = {2014},
    pages = {1739--1752},
}

@article{noe_variational_2013,
	title = {A variational approach to modeling slow processes in stochastic dynamical systems},
	volume = {11},
	doi = {10.1137/110858616},
	number = {2},
	journal = {Multiscale Modeling and Simulation},
	author = {Noé, Frank and Nüske, Feliks},
	year = {2013},
	pages = {635--655},
}

@article{zhang_reliable_2017,
    title = {Reliable {Approximation} of {Long} {Relaxation} {Timescales} in {Molecular} {Dynamics}},
    volume = {19},
    issn = {1099-4300},
    url = {https://www.mdpi.com/1099-4300/19/7/367},
    doi = {10.3390/e19070367},
    language = {en},
    number = {7},
    urldate = {2026-04-22},
    journal = {Entropy},
    author = {Zhang, Wei and Schütte, Christof},
    month = jul,
    year = {2017},
    pages = {367},
}

@article{zhang_solving_2022,
    title = {Solving eigenvalue {PDEs} of metastable diffusion processes using artificial neural networks},
    volume = {465},
    issn = {00219991},
    url = {https://linkinghub.elsevier.com/retrieve/pii/S0021999122004399},
    doi = {10.1016/j.jcp.2022.111377},
    language = {en},
    urldate = {2026-04-22},
    journal = {Journal of Computational Physics},
    author = {Zhang, Wei and Li, Tiejun and Schütte, Christof},
    month = sep,
    year = {2022},
    pages = {111377},
}

@article{nuske_efficient_2023,
    title = {Efficient approximation of molecular kinetics using random {Fourier} features},
    volume = {159},
    url = {https://doi.org/10.1063/5.0162619},
    number = {7},
    journal = {The Journal of Chemical Physics},
    author = {Nüske, Feliks and Klus, Stefan},
    year = {2023},
    pages = {074105},
}

@article{rahimiRandomFeaturesLargescale2007a,
    title = {Random features for large-scale kernel machines},
    volume = {20},
    journal = {Advances in Neural Information Processing Systems},
    author = {Rahimi, A. and Recht, B.},
    year = {2007},
}

@article{zhangEffectiveDynamicsGiven2016a,
    title = {Effective dynamics along given reaction coordinates, and reaction rate theory},
    volume = {195},
    journal = {Faraday discussions},
    author = {Zhang, Wei and Hartmann, Carsten and Schütte, Christof},
    year = {2016},
    pages = {365--394},
}

@article{nuske_spectral_2021,
    title = {Spectral properties of effective dynamics from conditional expectations},
    volume = {23},
    doi = {10.3390/e23020134},
    number = {2},
    journal = {Entropy},
    author = {Nüske, Feliks and Koltai, Péter and Boninsegna, Lorenzo and Clementi, Cecilia},
    year = {2021},
    pages = {1--25},
}

@article{nateghi_kinetically_2025,
    title = {Kinetically {Consistent} {Coarse} {Graining} {Using} {Kernel}-{Based} {Extended} {Dynamic} {Mode} {Decomposition}},
    volume = {21},
    copyright = {https://creativecommons.org/licenses/by/4.0/},
    issn = {1549-9618, 1549-9626},
    url = {https://pubs.acs.org/doi/10.1021/acs.jctc.5c00479},
    doi = {10.1021/acs.jctc.5c00479},
    language = {en},
    number = {15},
    urldate = {2025-10-08},
    journal = {Journal of Chemical Theory and Computation},
    author = {Nateghi, Vahid and Nüske, Feliks},
    month = aug,
    year = {2025},
    pages = {7236--7248},
}

@article{perez-hernandez_identification_2013,
    title = {Identification of slow molecular order parameters for {Markov} model construction},
    volume = {139},
    issn = {0021-9606, 1089-7690},
    doi = {10.1063/1.4811489},
    language = {en},
    number = {1},
    urldate = {2026-04-15},
    journal = {The Journal of Chemical Physics},
    author = {Pérez-Hernández, Guillermo and Paul, Fabian and Giorgino, Toni and De Fabritiis, Gianni and Noé, Frank},
    month = jul,
    year = {2013},
    pages = {015102},
}

@article{molgedey_separation_1994,
    title = {Separation of a mixture of independent signals using time delayed correlations},
    volume = {72},
    copyright = {http://link.aps.org/licenses/aps-default-license},
    issn = {0031-9007},
    url = {https://link.aps.org/doi/10.1103/PhysRevLett.72.3634},
    doi = {10.1103/PhysRevLett.72.3634},
    language = {en},
    number = {23},
    urldate = {2026-05-18},
    journal = {Physical Review Letters},
    author = {Molgedey, L. and Schuster, H. G.},
    month = jun,
    year = {1994},
    pages = {3634--3637},
}

@article{mardt_vampnets_2018,
    title = {{VAMPnets} for deep learning of molecular kinetics},
    volume = {9},
    issn = {2041-1723},
    url = {https://www.nature.com/articles/s41467-017-02388-1},
    doi = {10.1038/s41467-017-02388-1},
    language = {en},
    number = {1},
    urldate = {2026-05-18},
    journal = {Nature Communications},
    author = {Mardt, Andreas and Pasquali, Luca and Wu, Hao and Noé, Frank},
    month = jan,
    year = {2018},
    pages = {5},
}

@article{sprink_accurate_2026,
    title = {Accurate and robust analysis of molecular kinetics with random features},
    volume = {164},
    issn = {0021-9606, 1089-7690},
    url = {https://pubs.aip.org/jcp/article/164/11/114115/3384134/Accurate-and-robust-analysis-of-molecular-kinetics},
    doi = {10.1063/5.0317585},
    language = {en},
    number = {11},
    urldate = {2026-03-23},
    journal = {The Journal of Chemical Physics},
    author = {Sprink, Hauke and Zhu, Yanchen and Mey, Antonia S. J. S. and Nüske, Feliks},
    month = mar,
    year = {2026},
    pages = {114115},
}

@article{charron_navigating_2025,
    title = {Navigating protein landscapes with a machine-learned transferable coarse-grained model},
    volume = {17},
    issn = {1755-4330, 1755-4349},
    url = {https://www.nature.com/articles/s41557-025-01874-0},
    doi = {10.1038/s41557-025-01874-0},
    language = {en},
    number = {8},
    urldate = {2026-05-18},
    journal = {Nature Chemistry},
    author = {Charron, Nicholas E. and Bonneau, Klara and Pasos-Trejo, Aldo S. and Guljas, Andrea and Chen, Yaoyi and Musil, Félix and Venturin, Jacopo and Gusew, Daria and Zaporozhets, Iryna and Krämer, Andreas and Templeton, Clark and Kelkar, Atharva and Durumeric, Aleksander E. P. and Olsson, Simon and Pérez, Adrià and Majewski, Maciej and Husic, Brooke E. and Patel, Ankit and De Fabritiis, Gianni and Noé, Frank and Clementi, Cecilia},
    month = aug,
    year = {2025},
    pages = {1284--1292},
}

@book{pavliotis_stochastic_2014,
    address = {New York, NY},
    series = {Texts in {Applied} {Mathematics}},
    title = {Stochastic {Processes} and {Applications}: {Diffusion} {Processes}, the {Fokker}-{Planck} and {Langevin} {Equations}},
    isbn = {9781493913220 9781493913237},
    shorttitle = {Stochastic {Processes} and {Applications}},
    language = {eng},
    number = {60},
    publisher = {Springer},
    author = {Pavliotis, Grigorios A.},
    year = {2014},
}

@article{deuflhardRobustPerronCluster2005a,
    title = {Robust {Perron} cluster analysis in conformation dynamics},
    volume = {398},
    doi = {10.1016/j.laa.2004.10.026},
    journal = {Linear Algebra Appl},
    author = {Deuflhard, P. and Weber, M.},
    year = {2005},
    pages = {161--184},
}

@inproceedings{devergne_learning_2024,
    address = {Vancouver, BC, Canada},
    title = {Learning the {Infinitesimal} {Generator} of {Stochastic} {Diffusion} {Processes}},
    isbn = {9798331314385},
    url = {http://www.proceedings.com/079017-4377.html},
    doi = {10.52202/079017-4377},
    urldate = {2026-06-16},
    booktitle = {Advances in {Neural} {Information} {Processing} {Systems} 37},
    publisher = {Neural Information Processing Systems Foundation, Inc. (NeurIPS)},
    author = {Devergne, Timothée and Halconruy, Hélène and Kostic, Vladimir and Lounici, Karim and Pontil, Massimiliano},
    year = {2024},
    pages = {137806--137846},
}

@misc{kostic_laplace_2024,
    title = {Laplace {Transform} {Based} {Low}-{Complexity} {Learning} of {Continuous} {Markov} {Semigroups}},
    copyright = {Creative Commons Attribution 4.0 International},
    url = {https://arxiv.org/abs/2410.14477},
    doi = {10.48550/ARXIV.2410.14477},
    urldate = {2026-06-16},
    publisher = {arXiv},
    author = {Kostic, Vladimir R. and Lounici, Karim and Halconruy, Hélène and Devergne, Timothée and Novelli, Pietro and Pontil, Massimiliano},
    year = {2024},
}

@article{devergne_slow_2025,
    title = {Slow dynamical modes from static averages},
    volume = {162},
    issn = {0021-9606, 1089-7690},
    url = {https://pubs.aip.org/jcp/article/162/12/124108/3340487/Slow-dynamical-modes-from-static-averages},
    doi = {10.1063/5.0246248},
    language = {en},
    number = {12},
    urldate = {2026-06-16},
    journal = {The Journal of Chemical Physics},
    author = {Devergne, Timothée and Kostic, Vladimir and Pontil, Massimiliano and Parrinello, Michele},
    month = mar,
    year = {2025},
    pages = {124108},
}

@article{strand_adaptive_2026,
    title = {Adaptive {Tensor} {Train} {Metadynamics} for {High}-{Dimensional} {Free} {Energy} {Exploration}},
    volume = {22},
    copyright = {https://doi.org/10.15223/policy-029},
    issn = {1549-9618, 1549-9626},
    url = {https://pubs.acs.org/doi/10.1021/acs.jctc.6c00487},
    doi = {10.1021/acs.jctc.6c00487},
    language = {en},
    number = {11},
    urldate = {2026-06-16},
    journal = {Journal of Chemical Theory and Computation},
    author = {Strand, Nils E. and Yang, Siyao and Khoo, Yuehaw and Dinner, Aaron R.},
    month = jun,
    year = {2026},
    pages = {5420--5434},
}

@article{kazeev_low-rank_2012,
    title = {Low-{Rank} {Explicit} {QTT} {Representation} of the {Laplace} {Operator} and {Its} {Inverse}},
    volume = {33},
    issn = {0895-4798, 1095-7162},
    url = {http://epubs.siam.org/doi/10.1137/100820479},
    doi = {10.1137/100820479},
    language = {en},
    number = {3},
    urldate = {2026-06-16},
    journal = {SIAM Journal on Matrix Analysis and Applications},
    author = {Kazeev, Vladimir A. and Khoromskij, Boris N.},
    month = jan,
    year = {2012},
    pages = {742--758},
}

@article{ostlund_thermodynamic_1995,
    title = {Thermodynamic {Limit} of {Density} {Matrix} {Renormalization}},
    volume = {75},
    copyright = {http://link.aps.org/licenses/aps-default-license},
    issn = {0031-9007, 1079-7114},
    url = {https://link.aps.org/doi/10.1103/PhysRevLett.75.3537},
    doi = {10.1103/PhysRevLett.75.3537},
    language = {en},
    number = {19},
    urldate = {2026-06-16},
    journal = {Physical Review Letters},
    author = {Östlund, Stellan and Rommer, Stefan},
    month = nov,
    year = {1995},
    pages = {3537--3540},
}

@article{lindorff-larsen_how_2011,
    title = {How {Fast}-{Folding} {Proteins} {Fold}},
    volume = {334},
    issn = {0036-8075, 1095-9203},
    url = {https://www.science.org/doi/10.1126/science.1208351},
    doi = {10.1126/science.1208351},
    language = {en},
    number = {6055},
    urldate = {2026-06-18},
    journal = {Science},
    author = {Lindorff-Larsen, Kresten and Piana, Stefano and Dror, Ron O. and Shaw, David E.},
    month = oct,
    year = {2011},
    pages = {517--520},
}

@book{frenkelUnderstandingMolecularSimulation2023,
    address = {S.l.},
    edition = {Third edition},
    title = {Understanding molecular simulation: from algorithms to applications},
    isbn = {978-0-323-91318-8},
    shorttitle = {Understanding molecular simulation},
    language = {eng},
    publisher = {ELSEVIER ACADEMIC PRESS},
    author = {Frenkel, Daan and Smit, Berend},
    year = {2023},
}

@book{tuckermanStatisticalMechanicsTheory2023a,
    address = {Oxford},
    edition = {2nd ed},
    series = {Oxford graduate texts},
    title = {Statistical mechanics: theory and molecular simulation},
    isbn = {978-0-19-882556-2},
    shorttitle = {Statistical mechanics},
    language = {eng},
    publisher = {Oxford university press},
    author = {Tuckerman, Mark E.},
    year = {2023},
}

@book{ulam_collection_1960,
    title = {A collection of mathematical problems},
    publisher = {Interscience},
    author = {Ulam, S. M.},
    year = {1960},
}

@article{dellnitz_approximation_1999,
    title = {On the approximation of complicated dynamical behavior},
    volume = {36},
    doi = {10.1137/S0036142996313002},
    number = {2},
    journal = {SIAM J. Numer. Anal.},
    author = {Dellnitz, M. and Junge, O.},
    year = {1999},
    pages = {491--515},
}

@article{schutte_direct_1999,
    title = {A direct approach to conformational dynamics based on hybrid {Monte} {Carlo}},
    volume = {151},
    doi = {10.1006/jcph.1999.6231},
    number = {1},
    journal = {J. Comput. Phys.},
    author = {Schütte, C. and Fischer, A. and Huisinga, W. and Deuflhard, P.},
    year = {1999},
    pages = {146--168},
}

@article{noeConstructingEquilibriumEnsemble2009,
    title = {Constructing the equilibrium ensemble of folding pathways from short off-equilibrium simulations},
    volume = {106},
    issn = {0027-8424, 1091-6490},
    url = {https://pnas.org/doi/full/10.1073/pnas.0905466106},
    doi = {10.1073/pnas.0905466106},
    language = {en},
    number = {45},
    urldate = {2025-11-11},
    journal = {Proceedings of the National Academy of Sciences},
    author = {Noé, Frank and Schütte, Christof and Vanden-Eijnden, Eric and Reich, Lothar and Weikl, Thomas R.},
    month = nov,
    year = {2009},
    pages = {19011--19016},
}

@article{plattnerCompleteProteinProtein2017b,
    title = {Complete protein–protein association kinetics in atomic detail revealed by molecular dynamics simulations and {Markov} modelling},
    volume = {9},
    issn = {1755-4330},
    doi = {10.1038/nchem.2785},
    number = {10},
    journal = {Nature Chemistry},
    author = {Plattner, Nuria and Doerr, Stefan and Fabritiis, Gianni De and Noé, Frank},
    month = oct,
    year = {2017},
    pages = {1005--1011},
}

@article{knoverekOpeningCrypticPocket2021,
    title = {Opening of a cryptic pocket in beta-lactamase increases penicillinase activity},
    volume = {118},
    issn = {0027-8424, 1091-6490},
    url = {https://pnas.org/doi/full/10.1073/pnas.2106473118},
    doi = {10.1073/pnas.2106473118},
    language = {en},
    number = {47},
    urldate = {2025-11-13},
    journal = {Proceedings of the National Academy of Sciences},
    author = {Knoverek, Catherine R. and Mallimadugula, Upasana L. and Singh, Sukrit and Rennella, Enrico and Frederick, Thomas E. and Yuwen, Tairan and Raavicharla, Shreya and Kay, Lewis E. and Bowman, Gregory R.},
    month = nov,
    year = {2021},
    pages = {e2106473118},
}

@article{wuVariationalApproachLearning2020c,
    title = {Variational approach for learning {Markov} processes from time series data},
    volume = {30},
    number = {1},
    journal = {Journal of Nonlinear Science},
    author = {Wu, Hao and Noé, Frank},
    year = {2020},
    pages = {23--66},
}

@article{koopmanHamiltonianSystemsTransformation1931b,
    title = {Hamiltonian systems and transformation in {Hilbert} space},
    volume = {17},
    doi = {10.1073/pnas.17.5.315},
    number = {5},
    journal = {Proc. Natl. Acad. Sci. U S A},
    author = {Koopman, B. O.},
    year = {1931},
    pages = {315},
}

@article{mezicSpectralPropertiesDynamical2005a,
    title = {Spectral {Properties} of {Dynamical} {Systems}, {Model} {Reduction} and {Decompositions}},
    volume = {41},
    issn = {0924-090X},
    doi = {10.1007/s11071-005-2824-x},
    number = {1-3},
    journal = {Nonlinear Dynamics},
    author = {Mezić, Igor},
    month = aug,
    year = {2005},
    pages = {309--325},
}

@article{budisicAppliedKoopmanism2012b,
    title = {Applied {Koopmanism}},
    volume = {22},
    doi = {10.1063/1.4772195},
    number = {4},
    journal = {Chaos: An Interdisciplinary Journal of Nonlinear Science},
    author = {Budišić, M. and Mohr, R. and Mezić, I.},
    year = {2012},
}

@book{mauroyKoopmanOperatorSystems2020c,
    title = {The {Koopman} {Operator} in {Systems} and {Control}},
    volume = {484},
    isbn = {978-3-030-35712-2},
    publisher = {Springer International Publishing},
    editor = {Mauroy, Alexandre and Mezić, Igor and Susuki, Yoshihiko},
    year = {2020},
    doi = {10.1007/978-3-030-35713-9},
}

@article{bruntonModernKoopmanTheory2022,
    title = {Modern {Koopman} {Theory} for {Dynamical} {Systems}},
    volume = {64},
    issn = {0036-1445, 1095-7200},
    url = {https://epubs.siam.org/doi/10.1137/21M1401243},
    doi = {10.1137/21M1401243},
    language = {en},
    number = {2},
    urldate = {2025-11-11},
    journal = {SIAM Review},
    author = {Brunton, Steven L. and Budišić, Marko and Kaiser, Eurika and Kutz, J. Nathan},
    month = may,
    year = {2022},
    pages = {229--340},
}

@incollection{colbrookChapter4Multiverse2024,
    series = {Handbook of {Numerical} {Analysis}},
    title = {Chapter 4 - {The} multiverse of dynamic mode decomposition algorithms},
    volume = {25},
    url = {https://www.sciencedirect.com/science/article/pii/S1570865924000048},
    booktitle = {Numerical {Analysis} {Meets} {Machine} {Learning}},
    publisher = {Elsevier},
    author = {Colbrook, Matthew J.},
    editor = {Mishra, Siddhartha and Townsend, Alex},
    year = {2024},
    doi = {10.1016/bs.hna.2024.05.004},
    pages = {127--230},
}

@article{williams_datadriven_2015,
    title = {A {Data}–{Driven} {Approximation} of the {Koopman} {Operator}: {Extending} {Dynamic} {Mode} {Decomposition}},
    volume = {25},
    issn = {0938-8974, 1432-1467},
    shorttitle = {A {Data}–{Driven} {Approximation} of the {Koopman} {Operator}},
    doi = {10.1007/s00332-015-9258-5},
    language = {en},
    number = {6},
    urldate = {2026-04-15},
    journal = {Journal of Nonlinear Science},
    author = {Williams, Matthew O. and Kevrekidis, Ioannis G. and Rowley, Clarence W.},
    month = dec,
    year = {2015},
    pages = {1307--1346},
}

@book{bowman_introduction_2014,
    title = {An {Introduction} to {Markov} {State} {Models} and {Their} {Application} to {Long} {Timescale} {Molecular} {Simulation}},
    volume = {797},
    isbn = {978-94-007-7605-0},
    publisher = {Springer Netherlands},
    editor = {Bowman, Gregory R. and Pande, Vijay S. and Noé, Frank},
    year = {2014},
    doi = {10.1007/978-94-007-7606-7},
}

@article{schwantesModelingMolecularKinetics2015b,
    title = {Modeling {Molecular} {Kinetics} with {tICA} and the {Kernel} {Trick}},
    volume = {11},
    doi = {10.1021/ct5007357},
    number = {2},
    journal = {Journal of Chemical Theory and Computation},
    author = {Schwantes, C. R. and Pande, V. S.},
    year = {2015},
    pages = {600--608},
}

@article{klusKernelbasedApproachMolecular2018b,
    title = {A kernel-based approach to molecular conformation analysis},
    volume = {149},
    issn = {0021-9606},
    doi = {10.1063/1.5063533},
    number = {24},
    journal = {The Journal of Chemical Physics},
    author = {Klus, Stefan and Bittracher, Andreas and Schuster, Ingmar and Schütte, Christof},
    month = dec,
    year = {2018},
}

@article{chenNonlinearDiscoverySlow2019b,
    title = {Nonlinear discovery of slow molecular modes using state-free reversible {VAMPnets}},
    volume = {150},
    issn = {0021-9606},
    doi = {10.1063/1.5092521},
    number = {21},
    journal = {The Journal of Chemical Physics},
    author = {Chen, Wei and Sidky, Hythem and Ferguson, Andrew L.},
    month = jun,
    year = {2019},
}

@article{nateghi_consistent_2026,
    title = {Consistent projection of {Langevin} dynamics: {Preserving} thermodynamics and kinetics in coarse-grained models},
    volume = {113},
    issn = {2470-0045, 2470-0053},
    shorttitle = {Consistent projection of {Langevin} dynamics},
    url = {https://link.aps.org/doi/10.1103/wckl-dz9d},
    doi = {10.1103/wckl-dz9d},
    language = {en},
    number = {5},
    urldate = {2026-06-19},
    journal = {Physical Review E},
    author = {Nateghi, Vahid and Neureither, Lara and Moqvist, Selma and Hartmann, Carsten and Olsson, Simon and Nüske, Feliks},
    month = may,
    year = {2026},
    pages = {055308},
}

@article{moqvist_thermodynamic_2025,
    title = {Thermodynamic {Interpolation}: {A} {Generative} {Approach} to {Molecular} {Thermodynamics} and {Kinetics}},
    volume = {21},
    copyright = {https://creativecommons.org/licenses/by/4.0/},
    issn = {1549-9618, 1549-9626},
    shorttitle = {Thermodynamic {Interpolation}},
    url = {https://pubs.acs.org/doi/10.1021/acs.jctc.4c01557},
    doi = {10.1021/acs.jctc.4c01557},
    language = {en},
    number = {5},
    urldate = {2025-03-22},
    journal = {Journal of Chemical Theory and Computation},
    author = {Moqvist, Selma and Chen, Weilong and Schreiner, Mathias and Nüske, Feliks and Olsson, Simon},
    month = mar,
    year = {2025},
    pages = {2535--2545},
}

@article{donati_estimation_2018,
    title = {Estimation of the infinitesimal generator by square-root approximation},
    volume = {30},
    issn = {0953-8984, 1361-648X},
    url = {https://iopscience.iop.org/article/10.1088/1361-648X/aadfc8},
    doi = {10.1088/1361-648X/aadfc8},
    number = {42},
    urldate = {2026-06-19},
    journal = {Journal of Physics: Condensed Matter},
    author = {Donati, Luca and Heida, Martin and Keller, Bettina G and Weber, Marcus},
    month = oct,
    year = {2018},
    pages = {425201},
}

@inproceedings{meanti_estimating_2023,
    title = {Estimating {Koopman} operators with sketching to provably learn large scale dynamical systems},
    volume = {36},
    url = {https://proceedings.neurips.cc/paper_files/paper/2023/file/f3d1e34a15c0af0954ae36a7f811c754-Paper-Conference.pdf},
    booktitle = {Advances in {Neural} {Information} {Processing} {Systems}},
    publisher = {Curran Associates, Inc.},
    author = {Meanti, Giacomo and Chatalic, Antoine and Kostic, Vladimir and Novelli, Pietro and Pontil, Massimiliano and Rosasco, Lorenzo},
    editor = {Oh, A. and Naumann, T. and Globerson, A. and Saenko, K. and Hardt, M. and Levine, S.},
    year = {2023},
    pages = {77242--77276},
}

@article{aristoff_fast_2024,
    title = {The fast committor machine: {Interpretable} prediction with kernels},
    volume = {161},
    issn = {0021-9606, 1089-7690},
    shorttitle = {The fast committor machine},
    url = {https://pubs.aip.org/jcp/article/161/8/084113/3309968/The-fast-committor-machine-Interpretable},
    doi = {10.1063/5.0222798},
    language = {en},
    number = {8},
    urldate = {2026-07-06},
    journal = {The Journal of Chemical Physics},
    author = {Aristoff, David and Johnson, Mats and Simpson, Gideon and Webber, Robert J.},
    month = aug,
    year = {2024},
    pages = {084113},
}

@book{KhoK2018,
url = {https://doi.org/10.1515/9783110365832},
title = {Tensor Numerical Methods in Quantum Chemistry},
author = {Venera Khoromskaia and Boris N. Khoromskij},
publisher = {De Gruyter},
address = {Berlin, Boston},
doi = {doi:10.1515/9783110365832},
isbn = {9783110365832},
year = {2018},
lastchecked = {2026-07-16}
}

@book{Kho2018,
url = {https://doi.org/10.1515/9783110365917},
title = {Tensor Numerical Methods in Scientific Computing},
author = {Boris N. Khoromskij},
publisher = {De Gruyter},
address = {Berlin, Boston},
doi = {doi:10.1515/9783110365917},
isbn = {9783110365917},
year = {2018},
lastchecked = {2026-07-16}
}

@book{BalK2025, 
  place={Cambridge}, 
  title={Tensor Decompositions for Data Science}, publisher={Cambridge University Press}, 
  author={Ballard, Grey and Kolda, Tamara G.}, 
  year={2025}
}
%
%
%
%

\AddToHook{enddocument/afteraux}{%
\immediate\write18{
cp output.aux tt_gedmd.aux
}%
}
\end{document}